\documentclass{article}  
\usepackage[utf8]{inputenc}
\usepackage[margin=1in,top=15 mm]{geometry}
\usepackage{authblk}
\usepackage{setspace}
\usepackage[colorlinks,citecolor=blue,urlcolor=blue,bookmarks=false,hypertexnames=true]{hyperref} 
\usepackage{graphicx}
\usepackage{multirow}
\usepackage[numbers,sort&compress]{natbib}
\bibpunct{(}{)}{,}{n}{}{,}
\usepackage{verbatim}
\usepackage{float}
\usepackage{amsmath}
\usepackage{hyperref}
\usepackage[nameinlink,noabbrev]{cleveref}
\usepackage{csquotes}

\begin{document}

\sloppy

\title{\textbf{Enduring mechanical memory from the constitutive response of elastically recoverable nanostructured materials}}

\author[1,=]{Abhishek Gupta}
\author[1,=]{Bhanugoban Maheswaran}
\author[1]{Nicholas Jaegersberg}
\author[1]{Komal Chawla}
\author[1,*]{Ramathasan Thevamaran}

\affil[1]{Department of Mechanical Engineering, University of Wisconsin-Madison, Madison, WI, 53706, USA}

\affil[=]{Equally Contributed Authors}

\affil[*]{Corresponding author: thevamaran@wisc.edu}

\maketitle

\vspace*{-\baselineskip}
\vspace*{-\baselineskip}
\par\noindent\rule{\textwidth}{0.5pt}
\smallskip

\onehalfspacing

\begin{abstract}

Mechanical memory and computing are gaining significant traction as means to augment traditional electronics for robust and energy efficient performance in extreme environments. However, progress has largely focused on bistable metamaterials, while traditional constitutive memory effects have been largely overlooked—primarily due to the absence of compelling experimental demonstrations in elastically recoverable materials. Here, we report constitutive return point memory (RPM) in elastically recoverable, vertically aligned carbon nanotube (VACNT) foams, analogous to magnetic hysteresis-based RPM utilized in hard drives. Unlike viscoelastic fading memory, VACNTs exhibit non-volatile memory arising from rate-independent nanoscale friction. We find that the interplay between RPM and frictional dissipation enables independent tunability of the VACNTs’ dynamic modulus, allowing for both on-demand softening and stiffening. We leverage this property to experimentally demonstrate tunable wave speed in a VACNT array with rigid interlayers, paving the way for novel shock limiters, elastodynamic lensing, and wave-based analog mechanical computing.


\end{abstract}

\onehalfspacing

\doublespacing

\newpage

\section*{Introduction}

The ability to encode, retrieve, and erase information by leveraging a material's constitutive response, known as the material-memory effect, manifests in diverse forms across various material systems \cite{keim2019memory,paulsen2024mechanical}. Distinct from the broadly studied metamaterials-based digital memory which rely on carefully engineered elastic-energy landscapes through multi-stable geometric designs \cite{yasuda2021mechanical,chen2021reprogrammable,raney2016stable,he2024programmable,mei2021mechanical,shohat2022memory,li2024reprogrammable}, the material-memory effect arises from the constitutive response, reflecting the material's remembrance of the history of applied stimuli. In such cases, simply understanding and exploiting the constitutive response is sufficient for tunable functionalities---no complex metamaterial design is required. This approach is analogous to examples of constitutive-based memory observed in current-voltage hysteresis in memory resistor (memristor) based devices \cite{strukov2008missing}, shape memory triggered by various physical stimuli \cite{sullivan2018hydration,buehler1963effect}, memory of previously applied maximum stress observed in the Mullins effect \cite{mullins1969softening}, and return point memory (RPM) in magnetic hysteresis for data storage in hard drives and magnetic tapes \cite{perkovic1997improved,deutsch2004return}.
In ferromagnetic materials exhibiting magnetic hysteresis, RPM refers to the ability to retain memory of the switching point of the external magnetic field's direction, as the hysteresis curve returns to this point when the field's direction is switched back \cite{perkovic1997improved}.
Similarly, in mechanics, RPM has been observed in the stress-strain response of elasto-plastic materials \cite{IWan}. After loading the material into the plastic regime and allowing it to unload, it retains an imprint of the unloading point, enabling it to retrace the path upon reloading and return to the original monotonous stress-strain curve \cite{magerle2024rate,IWan,paulsen2024mechanical}. However, unlike the RPM in ferromagnetism, the stress-strain RPM in elasto-plastic materials, leads to permanent alteration of the material through plasticity. Until now, RPM has not been observed in elastically recoverable materials, leaving their potential for practical applications untapped.

Here, we demonstrate highly consistent and repeatable RPM in elastically recoverable vertically aligned carbon nanotube (VACNT) foams synthesized by a scalable chemical vapor deposition (CVD) process. VACNT foams are renowned for their high specific energy absorption \cite{cao2005super}, fatigue resistance \cite{suhr2007fatigue}, and thermal stability \cite{xu2010carbon,yang2011modeling}. While their mechanical and thermal properties have been extensively studied \cite{chawla2024superior}, little is known about their hysteretic stress-strain response under non-monotonic loading, such as partial cyclic loading \cite{sethna1993hysteresis,ortin1992preisach}, which is crucial for understanding the constitutive memory effects identified in this manuscript. Moreover, their dynamic mechanical behavior at high frequencies and across various strain amplitudes remains elusive. In this study, we comprehensively characterize the mechanical behavior of VACNT foams as a function of strain rate and strain magnitude through stress-relaxation, quasistatic cyclic compression, and dynamic mechanical analysis (DMA) experiments. We found that unlike viscoelastic polymeric foams, VACNT foams do not exhibit stress relaxation and demonstrate strain-rate-independent behavior. Contrary to previous speculations \cite{xu2010carbon,suhr2007fatigue,lattanzi2012nonlinear}, we provide evidence for a non-viscoelastic mechanism of energy dissipation in VACNT foams, governed by frictional interactions between nanotubes. We developed a novel deformation-dependent stick-slip frictional (DDSSF) model, consisting of springs and frictional sliders \cite{muravskii2004frequency}, to explain the rate-independent behavior. Our model accurately captures the RPM in VACNT foams and provides insights into its origin, attributed to stick-slip frictional interactions among CNTs \cite{coveney1995triboelastic,IWan}. We show that RPM and deformation-dependent friction lead to dual tunability in the dynamic modulus of VACNT foams: strain amplitude-dependent softening, known as the Payne effect \cite{payne1964elasticity}, and static precompression-dependent stiffening. We experimentally demonstrate that these effects enable slowing of traveling stress pulses with increasing impulse amplitude, yet speed up under greater static pre-compression in a multilayer array of VACNT foams with elastic aluminum interlayers. This dual tunability facilitates amplitude-programmable wave propagation crucial for various wave lensing and diffusing applications \cite{spadoni2010generation}. Our work expands the scope of material-memory to address long-standing engineering challenges and establishes RPM as a novel addition to the growing field of mechanical memory, alongside metamaterials.

\section*{Results}

\subsection*{Fading memory and Return point memory (RPM)}

\begin{figure}[!ht]
	\centering
	\includegraphics[width=\textwidth]{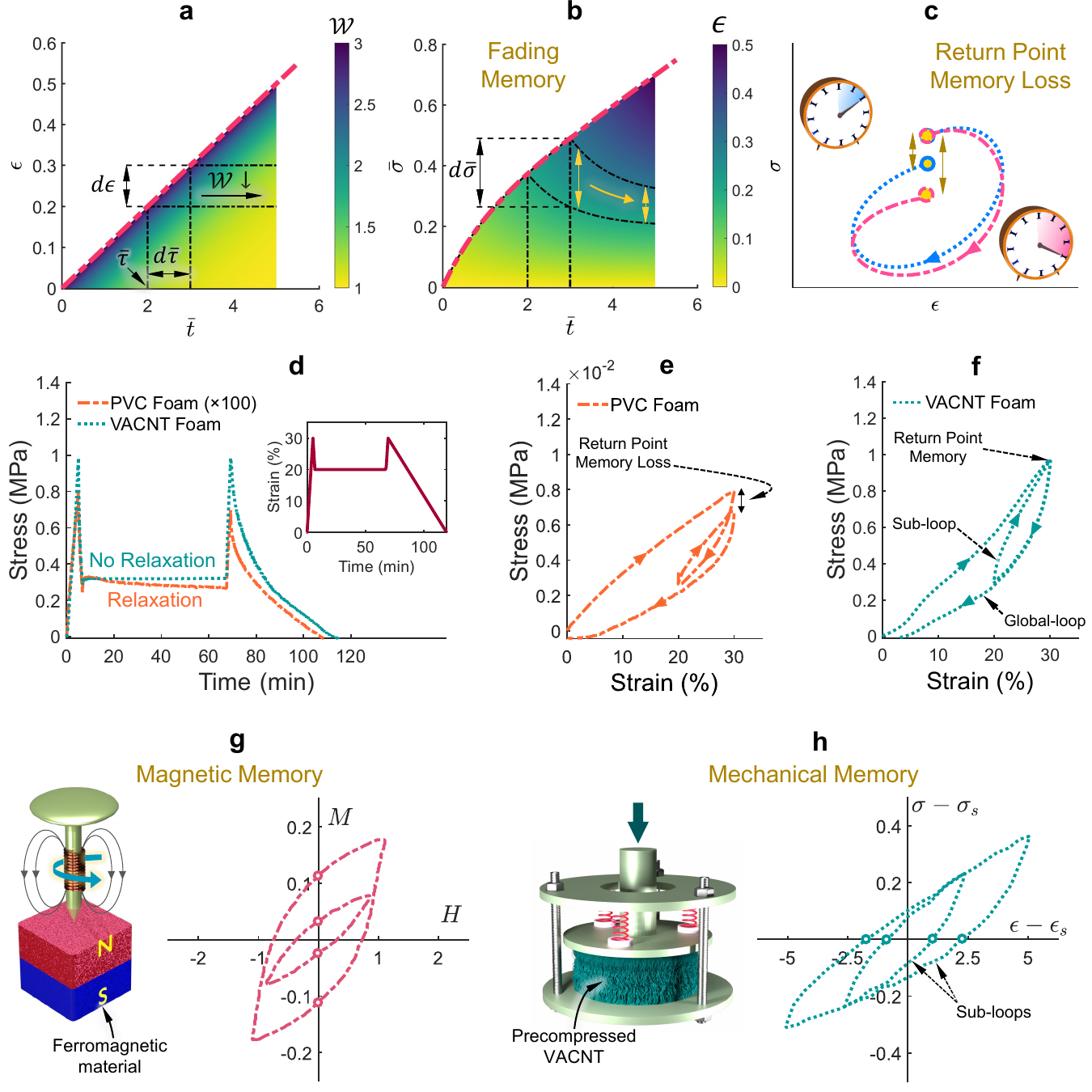}
	\caption{\textbf{Fading memory and return point memory.} (a) Plot of strain applied on a SLS material as a function of dimensionless time. A small increment in strain $d\epsilon$ applied from $\bar{\tau}$ to $\bar{\tau}+d\bar{\tau}$ is shown. The weight of each strain increment as a function of time is represented using a colormap (see details in \hyperref[section:sd2]{SI}). (b) Resultant dimensionless stress as a function of dimensionless time with contribution of each strain increment to the stress is shown using colormap. (c) An illustration of return point memory loss for different time intervals. (d) Comparison of stress as a function of time for VACNT foam and viscoelastic PVC foam, corresponding to the applied strain shown in the inset. PVC foam's stress is multiplied by 100 for better comparison. (e) Return point memory loss in viscoelastic PVC foam. (f) Return point memory in VACNT foam. (g) Illustration of magnetic memory achieved via ferromagnetic hysteresis \cite{sethna1993hysteresis}. (h) Illustration of mechanical memory through stress-strain hysteresis sub-loop in a precompressed VACNT sample.}
	\label{fig1}
\end{figure}

The foremost requirement for any non-volatile memory is the indefinite stability of the encoded information. For example, the polarity of ferromagnetic domains established during magnetic hysteresis remains permanently aligned even after the external magnetic field is removed. In contrast, the mechanical stress-strain hysteresis observed in soft dissipative materials, such as elastomers and polymeric foams, does not retain a lasting memory of the applied stress. Instead, they display fading memory due to their viscoelastic behavior. Fading memory implies that the magnitude of the effect of a recent cause is significantly higher than that of an earlier cause of the same magnitude \cite{christensen1972restrictions,lakes2009viscoelastic}. To elaborate, consider a standard linear solid (SLS) viscoelastic material subjected to a strain that increases linearly with dimensionless-time $(\bar{t})$ (\Cref{fig1}(a)). Let $d\epsilon$ represent an infinitesimal increase in strain from $\bar{t}=\bar{\tau}$ to $\bar{t}=\bar{\tau}+d\bar{\tau}$ (\Cref{fig1}(a)). The corresponding increase in dimensionless-stress (\Cref{fig1}(b)) can be expressed as follows (see details in \hyperref[section:sd2]{SI})

\begin{equation}
    d\bar{\sigma} \downarrow = \mathcal{W}\left(\bar{t}-\bar{\tau}\right)\downarrow\times d\epsilon
    \label{btz}
\end{equation}

where $\mathcal{W}$ represents the weight associated with $d\epsilon$, such that $\mathcal{W}(\bar{t} < \bar{\tau}) = 0$, and the down arrow $(\downarrow)$ indicates temporal decay. The magnitude of $\mathcal{W}$ decreases with time due to relaxation (\Cref{fig1}(a)), resulting in a fading memory effect (\Cref{fig1}(b)). Alternatively, if the strain is increased in equal steps at equal time intervals, the stress increment $d\bar{\sigma}$ resulting from the most recent strain increment $d\epsilon$ will contribute more to the total stress ($\bar{\sigma}$), whereas the $d\bar{\sigma}$ associated with earlier $d\epsilon$ values will have faded more significantly (\Cref{fig1}(b)), as they have had more time to relax. RPM loss is observed during the cyclic loading of viscoelastic materials because the memory of the switching point, where the loading direction is reversed, fades over time. The longer the duration, the greater the fading effect, as illustrated in \Cref{fig1}(c). 

Unlike viscoelastic polymeric foams, the VACNT foams do not undergo stress relaxation, enabling the memory of loading-direction reversal to persist indefinitely, which leads to RPM. In \Cref{fig1}(d), we present the experimentally measured stress response of a VACNT foam sample alongside a viscoelastic polyvinyl chloride (PVC) foam subjected to a prescribed strain profile over time (inset of \Cref{fig1}(d)). Both materials were first compressed (loaded) to a certain maximum strain $(30\%)$, partially unloaded $(-10\%)$, and then held at a constant strain for an extended period. As shown in \Cref{fig1}(d), the PVC foam exhibits stress relaxation, whereas the stress in the VACNT foam remains constant, indicating the absence of relaxation. Upon reloading to the maximum strain $(30\%)$, the PVC foam fails to retrace the previous unloading point, resulting in RPM loss (\Cref{fig1}(e)). In contrast, the time-independent behavior of VACNT foam enables it to close the hysteresis sub-loop, exhibiting RPM (\Cref{fig1}(f)). To maintain consistency across the manuscript, we establish the following terminology: the hysteresis loop obtained from cyclic loading of an unstrained VACNT sample is referred to as the global-loop, while the hysteresis loop resulting from any cyclic perturbation applied to a statically precompressed VACNT sample is termed the sub-loop (\Cref{fig1}(f)).

\Cref{fig1}(f) presents the first demonstration of stress-strain RPM in a soft, elastically recoverable material. In \Cref{fig1}(g) and (h), we draw analogy between ferromagnetic memory and VACNT-based mechanical memory. In ferromagnetic hysteresis \cite{sethna1993hysteresis}, the remanent magnetization $(M)$ after the removal of an external magnetic field $(H=0)$ can exist in two different states $(\pm)$, representing bits. Similarly, the hysteretic sub-loop of a VACNT sample precompressed under a constant static stress can retain two stable strain values, representing mechanical bits. The absence of both stress relaxation and creep in VACNTs preserves the established state of stress and strain, ensuring stable mechanical bits for enduring material memory. Moreover, in magnetic hysteresis, the size of the hysteresis loop can be shrunk or expanded by varying the amplitude of the cyclic external magnetic field $(H)$\cite{sethna1993hysteresis}, allowing the remnant magnetization $(M)$ to continuously take any value along the y-axis as shown in \Cref{fig1}(g). Similarly, in VACNTs, the size of the hysteretic sub-loop can be shrunk or enlarged depending on the extent of strain partially unloaded (\Cref{fig1}(h)). This allows remnant strains to take continuous values along the x-axis in \Cref{fig1}(h) at a constant static stress. In contrast, metamaterials with geometric bistable states are limited to two discrete states: ON and OFF \cite{chen2021reprogrammable,yasuda2017origami}.  

To understand the fundamental origin of RPM, its influence on the dynamic mechanical properties of VACNT foams, and the potential applications enabled by the resulting tunable behavior, we perform comprehensive experimental characterization across a broad range of strains and strain rates.

\subsection*{Time-independent VACNT foams}

VACNT foams exhibit a porous morphology composed of entangled multi-walled CNTs that behave like slender elastic struts (\Cref{figs1}), undergoing bending and buckling under compression---similar to open-cell polymeric foams. Owing to their morphological similarity and hysteretic response, the energy dissipation in VACNT foams has been erroneously attributed to viscoelasticity in the literature \cite{xu2010carbon,suhr2007fatigue,lattanzi2012nonlinear}. While energy dissipation in polymeric foams primarily arises from viscoelasticity, dissipation in VACNT foams is thought to occur predominantly through interfacial friction between nanotubes \cite{suhr2005viscoelasticity,cao2005super}. Experimental observations of viscoelastic time-dependent behaviors such as stress relaxation and creep deformation in VACNTs likely result from either incomplete preconditioning of the sample \cite{suhr2007fatigue} or misinterpretation of strain overshoot from the load frame as relaxation of the VACNT foam \cite{lattanzi2012nonlinear}.

\begin{figure}[htp!]
	\centering
	\includegraphics[width=0.95\textwidth]{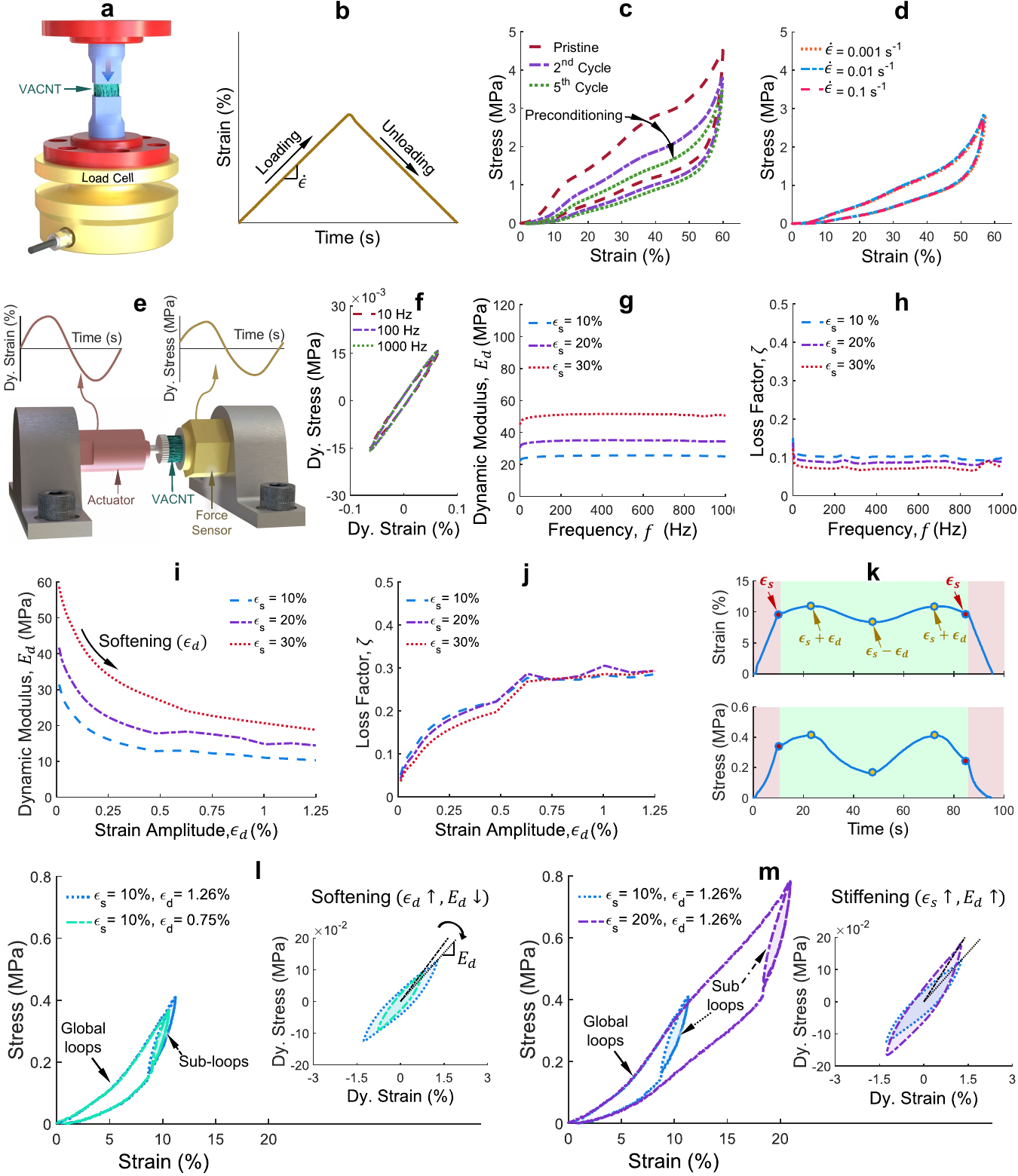}
	\caption{\textbf{Mechanical behavior of VACNT foams.} (a) Illustration of the Instron load frame used for quasistatic compression and stress-relaxation experiments. (b) Applied ramp strain with a constant strain-rate for quasistatic cyclic compression. (c) Quasistatic ramp-compression of a pristine VACNT sample undergoing mechanical preconditioning. (d) Preconditioned stress-strain global-hysteresis loops, showing rate-independency. (e) Illustration of our custom-built experimental apparatus for performing dynamic mechanical analysis. (f) Stress-strain hysteretic sub-loops for applied sinusoidal strain at different frequencies. (g,h) Dynamic modulus and loss factor as functions of frequency $(\epsilon_d=0.063\%)$. (i) Experimentally measured dynamic modulus and (j) loss factor as functions of dynamic strain amplitude $(\epsilon_d)$ and static precompression $(\epsilon_s)$. (k) Strain applied as a function of time (top) to measure the dynamic mechanical properties. Corresponding stress response as a function of time (bottom). (l) Comparison of complete hysteretic stress-strain responses from DMA testing for a fixed $\epsilon_s=10\%$ and two different $\epsilon_d$, with the corresponding hysteretic sub-loops are enlarged and compared in the Inset. (m) Comparison of complete hysteretic stress-strain responses for a fixed $\epsilon_d=1.26\%$ and two different $\epsilon_s$, with the corresponding hysteretic sub-loops shown in the Inset.}
	\label{fig2}
\end{figure}

Here, we experimentally demonstrate rate-independent mechanical behavior and a non-viscoelastic mechanism of energy dissipation that underpins the RPM effects in VACNT foams. We synthesize VACNT foams using a floating catalyst thermal chemical vapor deposition (tCVD) process (see Materials and Methods) resulting in MWCNTs that are nominally aligned vertically along the direction of growth (\Cref{figs1}(c)) \cite{gupta2022origins}. When compressed, pristine VACNT samples exhibit irreversible softening (\Cref{fig2}(c)) as a function of compression cycles---a phenomenon known as the preconditioning effect---which originates from the rearrangement of CNTs and permanent strain induced at the nanoscale \cite{gupta2022origins}. The material progressively becomes compliant during the first few cycles before the softening eventually ceases, and the sample reaches a preconditioned state after 4-5 cycles (\Cref{fig2}(c)). \Cref{figs1}(d) shows an SEM micrograph of a preconditioned sample, indicating that the initial vertical alignment of CNTs has been altered. Once preconditioned, the stress-strain response remains stable for thousands of cycles unless the previously applied maximum strain is exceeded \cite{cao2005super,suhr2007fatigue,gupta2022origins}. To ensure a consistent mechanical response across various experiments, we first preconditioned our sample by applying cyclic compression up to $60\%$ strain at a strain rate of $0.01\;s^{-1}$ for $5$ cycles (\Cref{fig2}(b,c)). 

\subsubsection{Strain rate independent constitutive behavior}

\Cref{fig2}(d) shows the cyclic quasistatic hysteretic stress–strain response of a preconditioned VACNT sample measured over strain rates spanning three orders of magnitude, with the maximum strain kept below the preconditioning strain $(60\%)$. The hysteretic global loop, comprising both loading and unloading curves, remains unaffected by varying strain rates, indicating rate-independent behavior. In contrast to the strong rate dependence observed in viscoelastic polymeric foams and elastomers \cite{lakes2009viscoelastic}, this rate independence in VACNT foams suggests that energy dissipation arises from non-viscoelastic mechanisms. Additionally, the rate-independent behavior is consistent with the absence of stress relaxation observed earlier (\Cref{fig1}(d)). Our detailed stress-relaxation experiments on VACNT foam further confirmed the absence of stress relaxation across all applied strain magnitudes (see details in \hyperref[section:sd2]{SI} and \Cref{figs7}).

\Cref{figs2}(a,b) shows the cyclic quasistatic global hysteresis loops obtained by ramping the strain to various maximum values ($\epsilon_{\text{max}}$). \Cref{figs2}(c) compares the unloading portions of the hysteresis loops, scaled by normalizing stress and strain by their respective maximum values. The scaled unloading curves overlap, suggesting the unloading curve scales self-similarly as a function of the maximum strain (or maximum stress) while retaining its overall shape. Notably, this scaling no longer holds in regimes where the loading curve becomes nonlinear---for example, near the onset of densification $(\geq49.64\%)$ (\Cref{fig2}(d)) or in the initial nonlinear region $(\leq10\%)$ caused by surface irregularities in the sample (\Cref{figs2}(b)). The scaling holds in the regime where the loading curve remains approximately linear, which informs our choice of the $\epsilon_{max}$ range considered in \Cref{figs2}. In a later section, we demonstrate that the self-similar scaling property of the unloading curve plays a crucial role in enabling tunable dynamic mechanical properties in VACNT foams.

\subsubsection{Frequency independent dynamic behavior}

While the quasistatic ramp compression and stress-relaxation experiments unambiguously indicate time-independent mechanical behavior, these experimental methods probe only quasistatic strain rates. For structured and hierarchical materials, it is often argued that the bulk behavior becomes rate dependent when the frequency of the external load matches the characteristic resonances of the micro/nanoscale constituent features that remain dormant at low strain rates. 

Previously, VACNT foams have been shown to exhibit frequency-independent dynamic modulus and loss-factor at low frequencies up to $100\;\textnormal{Hz}$ \cite{xu2010carbon}, however their broadband behavior has not been well characterized. Since commercial load frames have limitations on operating frequency, we built a custom dynamic mechanical analyzer (\Cref{fig2}(e)) consisting of a piezoelectric actuator and a dynamic force sensor capable of measuring the dynamic mechanical properties of soft materials up to $1000\;\textnormal{Hz}$ (see Materials and Methods). To perform dynamic mechanical analysis (DMA), we first apply a static precompression strain on the sample $(\epsilon_s)$ and then subject it to a sinusoidal strain of a desired amplitude $(\epsilon_d)$ and frequency $(f)$ using the piezoelectric actuator (\Cref{fig2}(e)). The total strain $(\epsilon_T)$ can be expressed as follows

\begin{equation}
    \epsilon_T(t)=\epsilon_s+\epsilon_d\sin(2\pi f t)
\end{equation}

In response, the force sensor measures the oscillatory component of the force, which we convert to stress by normalizing it by the cross-sectional area of the sample (see Materials and Methods). The oscillatory components of stress and strain (\Cref{fig2}(e)) forms a hysteretic subloop centered at the origin as shown in \Cref{fig2}(f). A frequency sweep from $1\;\mathrm{Hz}$ to $1000\;\mathrm{Hz}$ at constant $\epsilon_d = 0.063\%$ and $\epsilon_s = 10\%$ reveals that the hysteretic sub-loops remain unchanged with frequency (\Cref{fig2}(f)), indicating broadband frequency-independent behavior. From these loops, we extract two key dynamic mechanical properties of VACNT foams---dynamic modulus $(E_d)$ and loss factor $(\zeta)$---defined as follows (see details in \hyperref[section:sd2]{SI}):  

\begin{equation}
    E_d = {\frac{\sigma_d} {\epsilon_d}}\;\;,\;\;\zeta={\frac{E_l}{\sqrt{E_d^2-E_l^2}}}\;\;,\;\;E_l={\frac{W_{dis}}{\pi\times{\epsilon_d^2}}}
    \label{eqdyp}
\end{equation}

Here, $\sigma_d$ is the dynamic stress amplitude, $E_l$ represents the effective loss modulus, $\sqrt{E_d^2-E_l^2}$ denotes the effective storage modulus (see details in \hyperref[section:sd2]{SI}), and $W_{dis}$ is the energy dissipated per unit volume, calculated from the area inscribed in the hysteretic sub-loop. Both $E_d$ and $\zeta$ are frequency independent (\Cref{fig2}(g,h)), affirming rate-independent behavior and non-viscoelastic origin of energy dissipation. Unlike the strongly frequency-dependent behavior of viscoelastic foams, the broadband frequency-independent response observed here is uncommon among dissipative foams reported in the literature.

\subsection*{Dynamic softening and stiffening governed by the global-hysteresis loop's unloading curve}

To examine the effects of $\epsilon_d$ and $\epsilon_s$, we measured the dynamic mechanical properties as a function of $\epsilon_d$, ranging from $0.0125\%$ to $1.25\%$, for various values of $\epsilon_s \leq 30\%$, while keeping the total strain ($\epsilon_T$) well below the onset of densification $(49.64\%)$. The dynamic modulus ($E_d$) decreases steadily with increasing $\epsilon_d$ (\Cref{fig2}(i)) for each applied $\epsilon_s$, demonstrating strain-amplitude-dependent softening, commonly known as the Payne effect \cite{payne1964elasticity,coveney1995triboelastic}. In contrast, the loss factor increases with amplitude (\Cref{fig2}(j)) because $E_l$ (\Cref{eqdyp}) exhibits only a weak dependence on amplitude (\Cref{figs3}(a)), while $\sqrt{E_d^2 - E_l^2}$ decreases sharply with amplitude (\Cref{figs3}(b)). For a fixed $\epsilon_d$, $E_d$ increases with $\epsilon_s$ indicating stiffening, whereas the loss factor remains almost unchanged (\Cref{fig2}(j)).

To understand the contrasting effects of dynamic $(\epsilon_d)$ and static strain $(\epsilon_s)$ on the dynamic modulus $(E_d)$, we examine the complete stress–strain trajectory from the DMA experiments, which involve ramping the strain to $\epsilon_s$, holding it constant, and then superimposing a dynamic sinusoidal strain (\Cref{fig2}(k)). Since the force sensor in our custom DMA setup records only time-varying forces and the actuator’s piezo stack has a limited stroke, we conducted this experiment using an Instron Electropulse E3000 system (see Materials and Methods).
\Cref{fig2}(k) shows the applied strain waveform and resulting stress response as functions of time. 
Note that while the dynamic component of the applied strain follows the intended sinusoidal form, the corresponding stress response is quasi-sinusoidal—its maxima and minima align with those of the strain, but the curves diverge elsewhere. As a result, the hysteresis loop exhibits a biconvex shape with slope discontinuities, forming sharp corners when the direction of strain application is reversed (\Cref{fig2}(f)), in contrast to the smooth, elliptical loops typically observed in viscoelastic foams \cite{lakes2009viscoelastic}. This slope discontinuity upon strain reversal suggests that the underlying dissipative mechanism is direction-dependent---a phenomenon we examine in detail in a later section.

In \Cref{fig2}(l), we compare the full stress–strain response for $\epsilon_d = 0.75\%$ and $\epsilon_d = 1.26\%$, both at the same precompression strain $\epsilon_s = 10\%$.
For both $\epsilon_d$, the initial loading in the global-loop up to the maximum strain $(\epsilon_s + \epsilon_d)$ follows the loading curve observed in the quasistatic ramp compression (\Cref{figs2}). The sinusoidal part of the applied strain $(\epsilon_d \sin(2\pi ft))$ (\Cref{fig2}(k)) results in a small hysteretic sub-loop comprising of an unloading half-cycle ($2\epsilon_d$ strain unloaded from $\epsilon_s+\epsilon_d$ to $\epsilon_s-\epsilon_d$) and a reloading half-cycle (loading the strain back from $\epsilon_s-\epsilon_d$ to $\epsilon_s+\epsilon_d$) (\Cref{fig2}(l), \Cref{figs3}(c,d)). Due to rate independence, the unloading half-cycle traces the corresponding portion of the global loop’s unloading curve up to the extent of strain unloaded (\Cref{figs3}(d)). In contrast, the reloading half-cycle closely matches an inverted version of the unloading half-cycle, indicating that the hysteresis loop exhibits Masing behavior \cite{IWan} (\Cref{figs3}(d)).
The reloading half-cycle fully encloses the hysteretic sub-loop, achieving RPM. 

It is evident that the unloading and reloading segments of the hysteretic subloop are composed of a small portion of the global unloading curve near the maximum strain and its inverted counterpart, respectively. As $\epsilon_d$ increases, a longer portion of the curve being traversed (\Cref{figs3}(d)). Because the slope of the global unloading curve decreases from the maximum strain toward zero strain (\Cref{figs2}(b,c)), the hysteretic sub-loop exhibits a lower average slope and appears tilted clockwise (inset of \Cref{fig2}(l)), resulting in a lower dynamic modulus.
In contrast, for a fixed $\epsilon_d$, increasing $\epsilon_s$ causes the sub-loop to tilt counter-clockwise (inset of \Cref{fig2}(m)), indicating dynamic stiffening. Notably, this two-way dependence of the dynamic modulus arises solely from the global unloading curve, while the global loading curve is nearly linear and plays no significant role. This unique behavior is not seen in other material systems, where such properties predominantly originate from the loading response. Here, we theoretically demonstrate that these properties originate from the scaling behavior of the global unloading curve discussed earlier (\Cref{figs2}).

As discussed earlier, the quasistatic unloading curve scales as a function of the maximum compressive strain ($\epsilon_{max}$)(\Cref{figs2}). We assume for the scaled unloading curve (\Cref{figs2}(c)), the normalized stress $(\sigma/\sigma_{max})$ is a function of the normalized strain $f(\epsilon/\epsilon_{max})$ as follows:

\begin{equation}
    \sigma = \sigma_{max}\times f\left( \frac{\epsilon}{\epsilon_{max}} \right)
    \label{eqscal}
\end{equation}

Using the above equation, applying RPM for hysteresis, we derive the following approximate equation for $E_d$ as a function of $\epsilon_s$ and $\epsilon_d$ (see details in \hyperref[section:sd2]{SI}).

\begin{equation}
    E_d\approx  3.981 \times \left[f'(1)-\frac{\epsilon_d}{(\epsilon_s+\epsilon_d)} f''(1) \right]
    \label{Ed9}
\end{equation}

Here, $f'(1)$ and $f''(1)$ denote the first and second derivatives of $f(\epsilon/\epsilon_{max})$, evaluated at $\epsilon/\epsilon_{max}=1$. Since the scaled unloading curve is monotonically increasing and concave-up (\Cref{figs2}(c)), both derivatives are positive. As a result, the analytical expression predicts that $E_d$ decreases with increasing $\epsilon_d$ but increases with increasing $\epsilon_s$. The general nature of the derivation underlying \Cref{Ed9} makes this relationship applicable to any material that exhibits similar scaling of unloading curve.
The comprehensive mechanical characterization of VACNT foams revealed their rate-independent behavior and tunable dynamic mechanical properties, but left the following questions unresolved: What physical mechanism gives VACNT foams their Masing behavior and RPM? And why does their unloading curve scale? We address these questions through our rate-independent, deformation-dependent stick–slip friction (DDSSF) model.

\subsection*{Deformation-dependent stick-slip frictional (DDSSF) model}

\begin{figure}[!t]
	\centering
	\includegraphics[width=\textwidth]{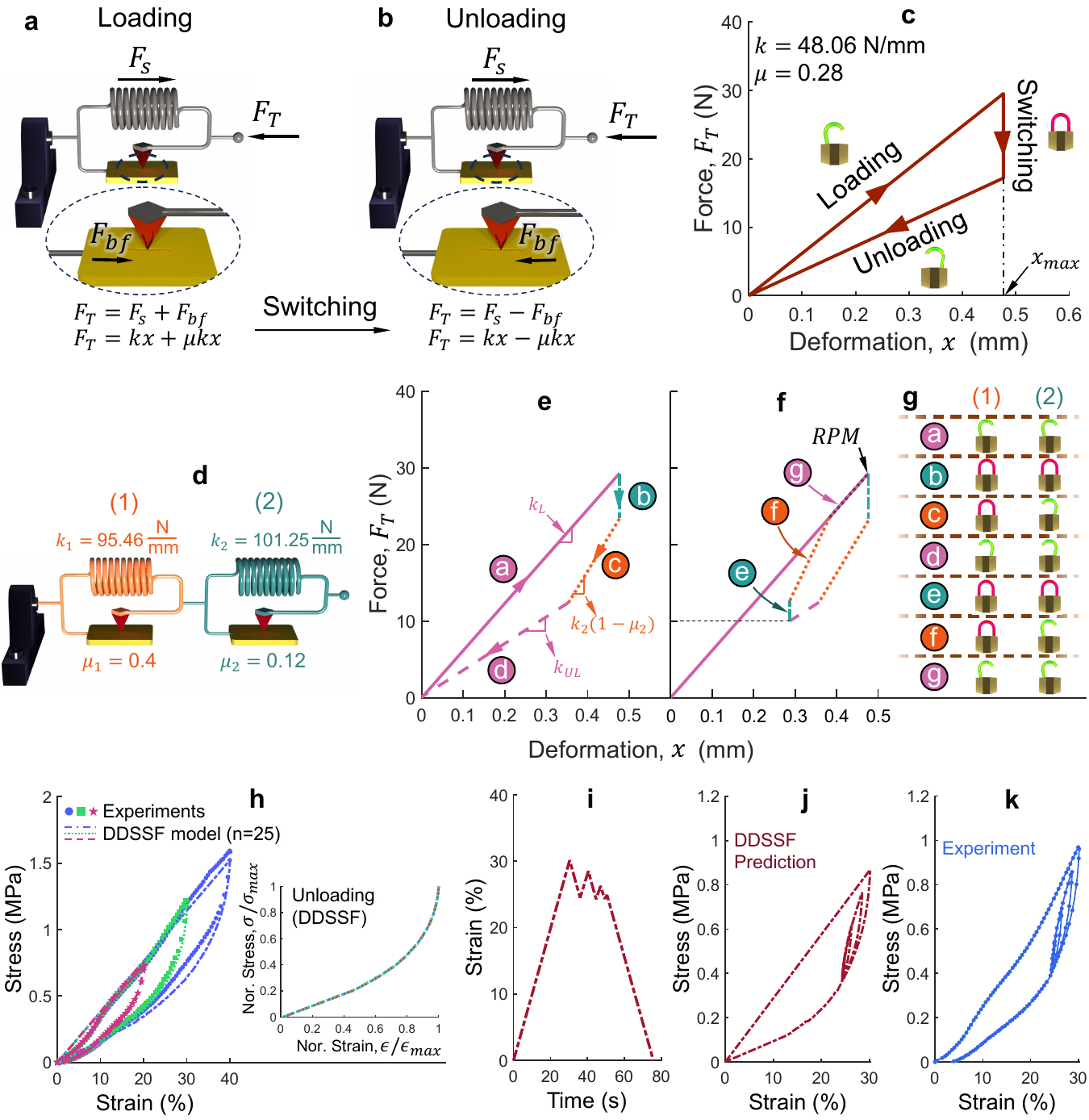}
 	\caption{\textbf{Deformation-dependent stick-slip frictional (DDSSF) model.} (a) A spring and a deformation-dependent Coulomb slider connected in parallel, undergoing compression (loading) and (b) unloading. (c) The hysteretic force-deformation cyclic response for a single spring-slider pair. (d) Two spring-slider pairs connected in series, each with different stiffness and friction coefficients. (e) The resulting cyclic force-deformation response. (f) Two spring-slider pairs exhibiting RPM. (g) Lock-unlock illustrations depicting the locking of Coulomb sliders during various phases of loading and unloading. (h) DDSSF model capturing experimental global hysteresis and scaling property of the unloading curve. (i) Strain applied as a function of time. (j) Stress-strain response predicted by the DDSSF model and (k) experimental verification.}
	\label{fig3}
\end{figure}

We propose a novel rate-independent frictional damping model that captures the experimentally observed mechanical behavior and provides insights into the origin of RPM and Masing behavior in VACNT foams. While viscoelastic models equipped with spring-dashpot elements have been widely used \cite{lion1998thixotropic}, their inherent rate dependence makes them unsuitable for VACNT foams. Moreover, the rate independence, Masing behavior, and slope discontinuities observed in VACNT foams suggest that a frictional damping model is appropriate, as these features are distinctive characteristics of such models. Instead of dashpots, frictional damping models use Coulomb sliders, typically illustrated as a pin sliding on a plate (\Cref{fig3}(a,b), \Cref{figs4}). The slider remains locked (stick) until the external force exceeds a break-free friction force ($F_{bf}$), triggering sliding (slip) motion \cite{IWan,coveney1995triboelastic,kamrin2014effect}. Frictional models, such as the Iwan model (\Cref{figs4}), have been shown to effectively capture unloading-point memory and the Masing behavior observed in elasto-plastic materials and bolted joints \cite{IWan}. Notably, these models are fundamentally analogous to the hysteron-based Preisach model used to describe ferromagnetic memory \cite{perkovic1997improved,sethna1993hysteresis}.

We modify the elasto-plastic frictional damping model to exhibit full elastic recovery by adopting Coulomb sliders with the break-free friction force ($F_{bf}$) proportional to the applied deformation ($x$)---a deformation-dependent stick-slip frictional (DDSSF) model. When connected in parallel with a spring ($F_s = kx$), the friction force in such a Coulomb slider ($F_{bf}$) opposes the relative sliding motion between the pin and the plate. As a result, it can either add to ($F_s + F_{bf}$) or subtract from ($F_s - F_{bf}$) the spring force, depending on the direction of sliding (\Cref{fig3}(a,b)).
The force in the spring ($F_s$) represents the elasticity of the CNT network undergoing bending and buckling, whereas $F_{bf}$ represents the internal friction between nanotubes. During compressive loading, the compaction of the VACNT foam enhances van der Waal frictional interactions \cite{grebenko2022local}, motivating our assumption that $F_{bf} \propto F_s$ \cite{kimball1927internal,muravskii2004frequency}. Since $F_s=kx$, the magnitude of friction force $F_{bf}$ becomes deformation-dependent $(F_{bf}\propto kx)$. As shown in \Cref{fig3}(a,b), an external force $F_T$ during loading (compression) and unloading (decompression) can be expressed as follows (see details in \hyperref[section:sd2]{SI})

\begin{equation}
    F_{bf}= \pm \mu F_s \;\;,\;\;F_s=kx
    \label{Fbf}
\end{equation}
\begin{equation}
    F_{T}=kx\times(1+\mu) \;\; \textnormal{(Loading)}, \;\;  F_{T}=kx \times(1-\mu) \;\; \textnormal{(Unloading)}
    \label{lounlo}
\end{equation}

where $\mu$ is the coefficient of friction $(0 \leq \mu < 1)$, and $k$ is the spring stiffness. Under cyclic compression, the hysteretic response of a single spring-slider pair (\Cref{fig3}(a,b)) consists of three segments: a linear loading branch, a linear unloading branch, and a switching segment (\Cref{fig3}(c)). When compressed (loaded) to a deformation of $x_{\text{max}}$, the spring force is $F_s = kx_{\text{max}}$, and the friction force is $F_{bf} = \mu k x_{\text{max}}$. When the pair is unloaded, the direction of the friction force in the slider switches from $+\mu k x_{\text{max}}$ to $-\mu k x_{\text{max}}$ to allow slipping in the opposite direction (\Cref{fig3}(a,b)). During this switching, the total force drops by $2\mu k x_{\text{max}}$ while the slider remains locked, resulting in no change in deformation $(x)$ (\Cref{fig3}(c)). Once the friction direction switches, the slider unlocks, and both the spring and the slider unload completely, following a linear force–deformation curve (\Cref{lounlo}). Unlike the Iwan model, which uses a constant $F_{bf}$ (\Cref{figs4}), our model results in no plastic deformation due to the deformation-dependent $F_{bf}$. Slope discontinuities occur at the transition points between loading and switching, and between switching and unloading, resembling those observed experimentally in VACNT foams.

Connecting multiple spring-slider pairs with varying stiffnesses and friction coefficients in series produces a gradual unloading response (\Cref{fig3}(d,e)), similar to that observed experimentally in VACNT foams. When an external compressive load is applied, all pairs compress simultaneously, following a linear loading curve (\Cref{fig3}(e)) whose slope corresponds to the effective loading stiffness $k_L$ (see details in \hyperref[section:sd2]{SI}). After loading to a certain maximum force $(F_{T,\text{max}})$, the sliders unlock sequentially in order of increasing $\mu$ during unloading. As shown in \Cref{fig3}(e,g), during segment (b) of unloading, while the force decreases, both pairs stay locked, and the deformation remains unchanged. Once the force drops sufficiently to unlock pair (2), the next unloading segment (b) follows a slope of $k_2(1 - \mu_2)$, while pair (1) stays locked, with its slider’s friction still in the process of switching direction. Once pair (1) also unlocks, both pairs unload simultaneously with a combined slope of $k_{UL}$ (see details in \hyperref[section:sd2]{SI}).

Using this same model of two spring-slider pairs, the fundamental origins of return-point memory (RPM) and Masing behavior can be understood. \Cref{fig3}(f) shows the hysteretic response when the model is partially unloaded to $F_T = 10\;\text{N}$, and then reloaded back to the previously applied maximum force, forming a hysteretic sub-loop. When the loading direction is reversed at $F_T = 10\;\text{N}$, the friction direction in the sliders, initially $-ve$, must switch back to $+ve$ to enable reloading, causing the sliders to unlock sequentially once again in the order of increasing $\mu$.
Once both pairs have unlocked, the reloading curve merges with the original global loading curve, ultimately exhibiting RPM. \Cref{fig3}(g) illustrates how the friction direction in the individual Coulomb sliders acts like mechanical bits, encoding the loading history. The evolution of friction direction during the sub-loop cycle reveals that RPM arises directly from both sliders resetting their friction direction to positive upon completing reloading. This reset allows the force ($F_T$) and deformation ($x$) to return to their previous magnitudes, thereby producing RPM. Furthermore, since the spring-slider pairs unlock in the same sequence each time the loading direction is reversed, an inverted symmetry emerges in the hysteretic sub-loop---characteristic of Masing behavior. However, for higher extents of partial unloading, this inverted symmetry gradually fades, and the Masing behavior ultimately ceases to exist, as detailed in \hyperref[section:sd2]{SI}.

Connecting many spring-slider pairs in series effectively captures the experimentally measured global hysteretic stress-strain response of VACNT foams (\Cref{fig3}(h), \Cref{figs5}(a,b)). We developed a MATLAB script (\hyperref[section:sd4]{GitHub} and \hyperref[section:sd2]{SI}) to compute the appropriate values of spring stiffnesses $(k_i,\;i=1,2,...n)$ and friction coefficients $(\mu_i)$ for a given number of spring-slider pairs $(n)$ to fit the experimental hysteretic response (\Cref{figs5}(b)). Once the fitted values of $k_i$ and $\mu_i$ are determined, the force–deformation response for any arbitrary force input can be calculated using another custom MATLAB script (see \hyperref[section:sd4]{GitHub} and \Cref{figs5}). Force and deformation are then normalized by the VACNT sample's cross-sectional area and thickness, respectively, to obtain stress and strain. \Cref{fig3}(h) shows the stress–strain response of a 25-pair DDSSF model accurately capturing the experimental hysteretic stress-strain response measured for various maximum strain levels. Inset of \Cref{fig3}(h) shows the DDSSF unloading curves also exhibit the scaling property previously observed in experiments (\Cref{figs2}). DDSSF model provides insight into the origin of this scaling: the friction force in each slider at the onset of unloading is proportional to the maximum applied force as $\mu_i F_{T,\text{max}} / (1 + \mu_i)$ (see details in \hyperref[section:sd2]{SI}). Since the friction force in all sliders scales uniformly with increasing $F_{T,\text{max}}$, the unloading curve also scales while preserving its shape.   

\Cref{fig3}(i,j,k) further demonstrates the accuracy of our model. We used the DDSSF model to predict the stress–strain response (\Cref{fig3}(j)) under an arbitrary strain input applied as a function of time (\Cref{fig3}(i)). The experimentally measured response (\Cref{fig3}(k)) closely matches the model prediction, validating the predictive capability of the DDSSF model. Additionally, the model captures the strain-amplitude-dependent softening and static-precompression-dependent stiffening behavior observed in DMA experiments. The hysteretic responses estimated using the DDSSF model, shown in \Cref{figs5}(c,d), closely match the experimental results (\Cref{fig2}(l,m)), further confirming the model’s accuracy. The compelling experimental evidence and theoretical verification we have presented demonstrate that our DDSSF model is the most accurate devised so far for VACNT foam materials in the literature \cite{fraternali2011multiscale}. This model can be further equipped with nonlinear springs to capture initial nonlinearity in the loading curve for $F_T<10\;N$ and nonlinearity in the loading curve for compressive strains beyond the onset of densification (\Cref{fig2}(i)), but this is beyond the scope of the present work.

The striking accuracy of the DDSSF model suggests a plausible mechanism for energy dissipation in VACNTs, wherein frictional interactions among MWCNTs span a range of length scales and magnitudes, which are being effectively represented by Coulomb sliders with varying friction coefficients. The model also proved successful in explaining the origin of RPM and how its interplay with deformation-dependent friction gives rise to a rich landscape of tunable dynamic mechanical properties, poised for novel engineering applications explored in the following section.

\subsection*{Elastic wave speed tunability in VACNT's RPM-enabled waveguides}

We design a periodic waveguide composed of VACNT foams alternately arranged with rigid aluminum interlayers to experimentally demonstrate tailored elastic wave speed by leveraging the independent dual-tunability of the VACNT's dynamic modulus as a function of dynamic strain amplitude $(\epsilon_d)$ and static strain $(\epsilon_s)$. The propagation speed of elastic waves in periodically-layered structures is governed by the dispersion relation, which depends on the constitutive and geometric properties of the layers \cite{hussein2014dynamics}. Nonlinearity in the constitutive response of those layers can modify the dispersion relation, causing the group velocity to change with amplitude \cite{narisetti2010perturbation,boechler2010discrete}. Such periodic waveguides featuring amplitude-dependent wave speeds can enable phenomena including wave delaying, focusing, defusing, selective amplitude blocking in acoustic lensing \cite{spadoni2010generation, bordiga2024automated}, and amplitude-selective filtering \cite{yang2012amplitude, mork2022nonlinear}.

\begin{figure}[!t]
	\centering
	\includegraphics[width=\textwidth]{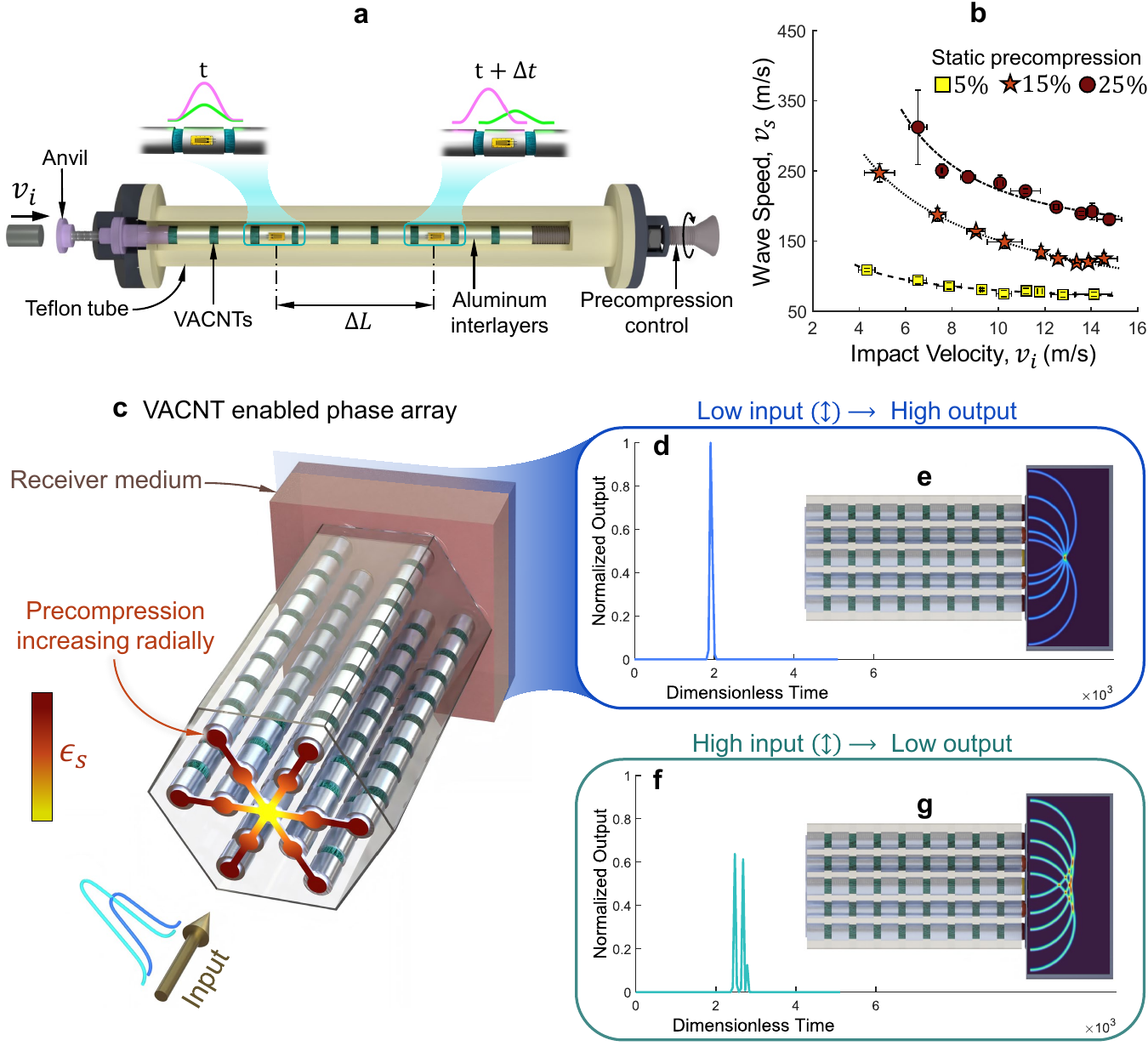}
	\caption{\textbf{Wave modulation using RPM-enabled waveguides.} (a) Illustration of the experimental setup showing a VACNT-Aluminum layered waveguide impacted by a projectile to generate traveling stress pulses of varying amplitude. (b) Speed of the traveling stress pulse plotted as a function of impact velocity $(v_i)$ for various static precompression levels. The error bars represent the standard deviation from three experimental repetitions. (c) Hexagonal phased array constructed with VACNT-Aluminum metastructure  with radially increasing precompression. (d,e) Wave coalescing occurring within the linear elastic receiver medium at low input design amplitudes. (f,g) The coalescing effect diminishes as the amplitude increases. In (d) and (f), the $x$-axis is normalized by $0.98\;\mathrm{ms}$, and the $y$-axis is normalized by $18.22\;\mathrm{MPa}$---the peak output amplitude for the low input case (see details in \hyperref[section:sd2]{SI}).}
	\label{fig4}
\end{figure}

\begin{figure}[H]
	\centering
	\includegraphics[width=\textwidth]{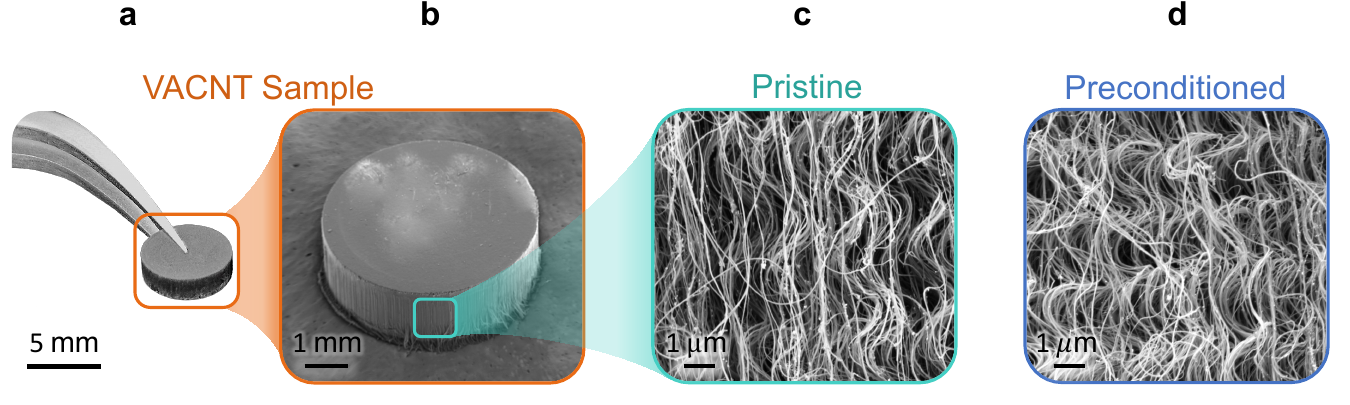}
	\caption{\textbf{Multi-lengthscale structural hierarchy of VACNT foam.} (a,b) VACNT sample cut from CVD synthesized VACNT film using biopsy punch. (c) Entangled morphology of CNTs at the microscale in a pristine sample. (d) Microscale morphology after preconditioning the sample showing vertical alignment is disturbed.}
	\label{figs1}
\end{figure}

\begin{figure}[H]
	\centering
	\includegraphics[width=\textwidth]{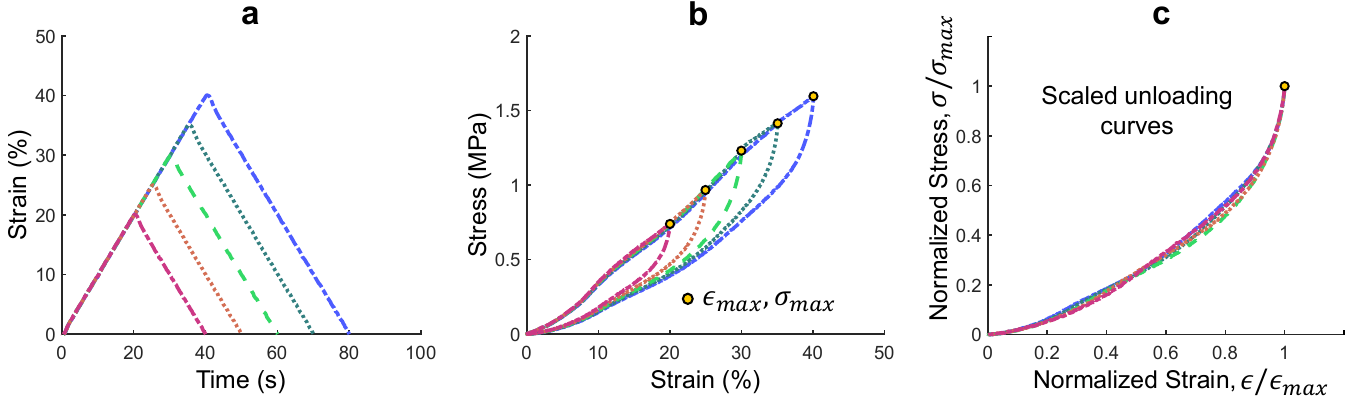}
	\caption{\textbf{Scaling of global unloading curves under quasistatic compression experiments.} (a) Strain applied as a ramp at a strain rate of $0.01\;s^{-1}$ up to various maximum strain levels, followed by unloading. (b) Resulting cyclic stress-strain responses plotted as a function of maximum strain. (c) Normalized unloading curves illustrating the observed scaling behavior. }
	\label{figs2}
\end{figure}

\begin{figure}[H]
	\centering
	\includegraphics[width=\textwidth]{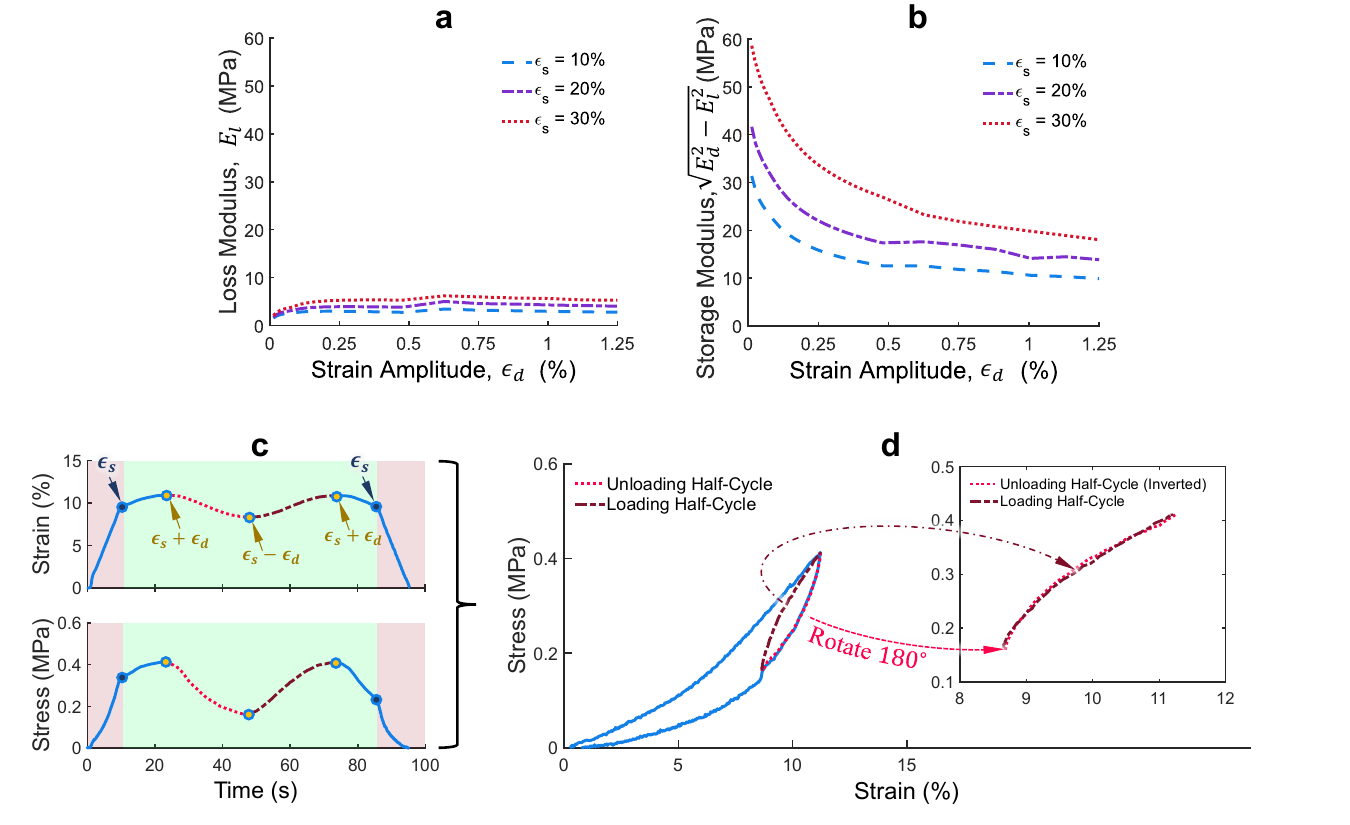}
	\caption{\textbf{Dynamic Mechanical Analysis (DMA).} (a) Loss modulus and (b) storage modulus as a function of dynamic strain amplitude for various precompression strains. (c) Strain and stress as a function of time. (d) Full stress-strain response involving ramping the strain to $10\%$ and applying dynamic sinusoidal strain. The resultant dynamic hysteretic response consisting of unloading half-cycle and loading half-cycle are shown via dotted curves. Loading half-cycle nearly coincides with the unloading half-cycle rotated $180^{\circ}$ (Inset). }
	\label{figs3}
\end{figure}

\begin{figure}[H]
	\centering
	\includegraphics[width=\textwidth]{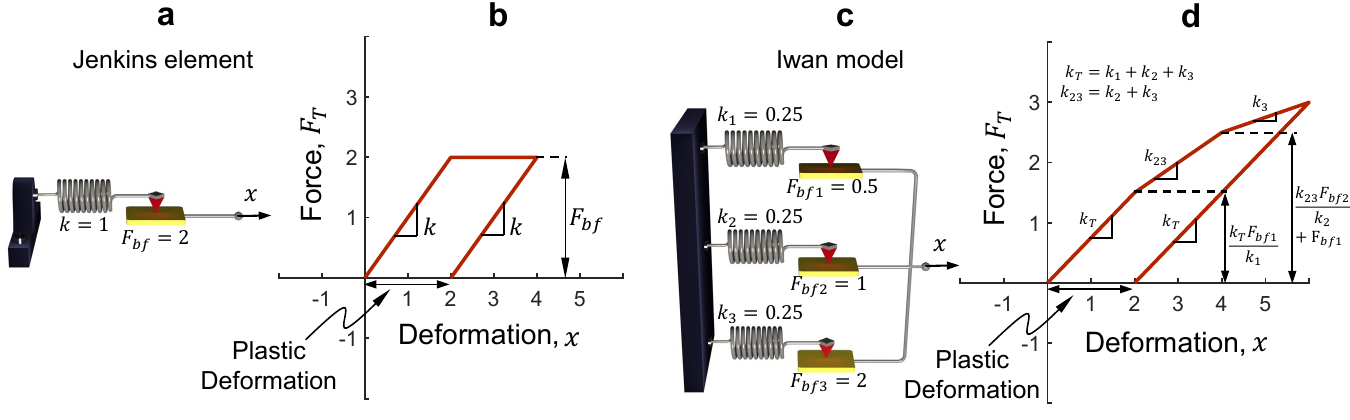}
	\caption{\textbf{Iwan model.} (a) Jenkins element---a spring and a Coulomb slider connected in series. (b) Force-deformation response of a Jenkins element with a spring stiffness of $k=1$ (arb. units) and a Coulomb slider break-free force of $F_{bf}=2$ (arb. units). (c) Multiple Jenkins elements connected in parallel or a parallel Iwan model. Each Jenkins element with a Coulomb slider having a different break-free force. (d) Force-deformation response of a parallel Iwan model, exhibiting plastic deformation.}
	\label{figs4}
\end{figure}

\begin{figure}[H]
	\centering
	\includegraphics[width=\textwidth]{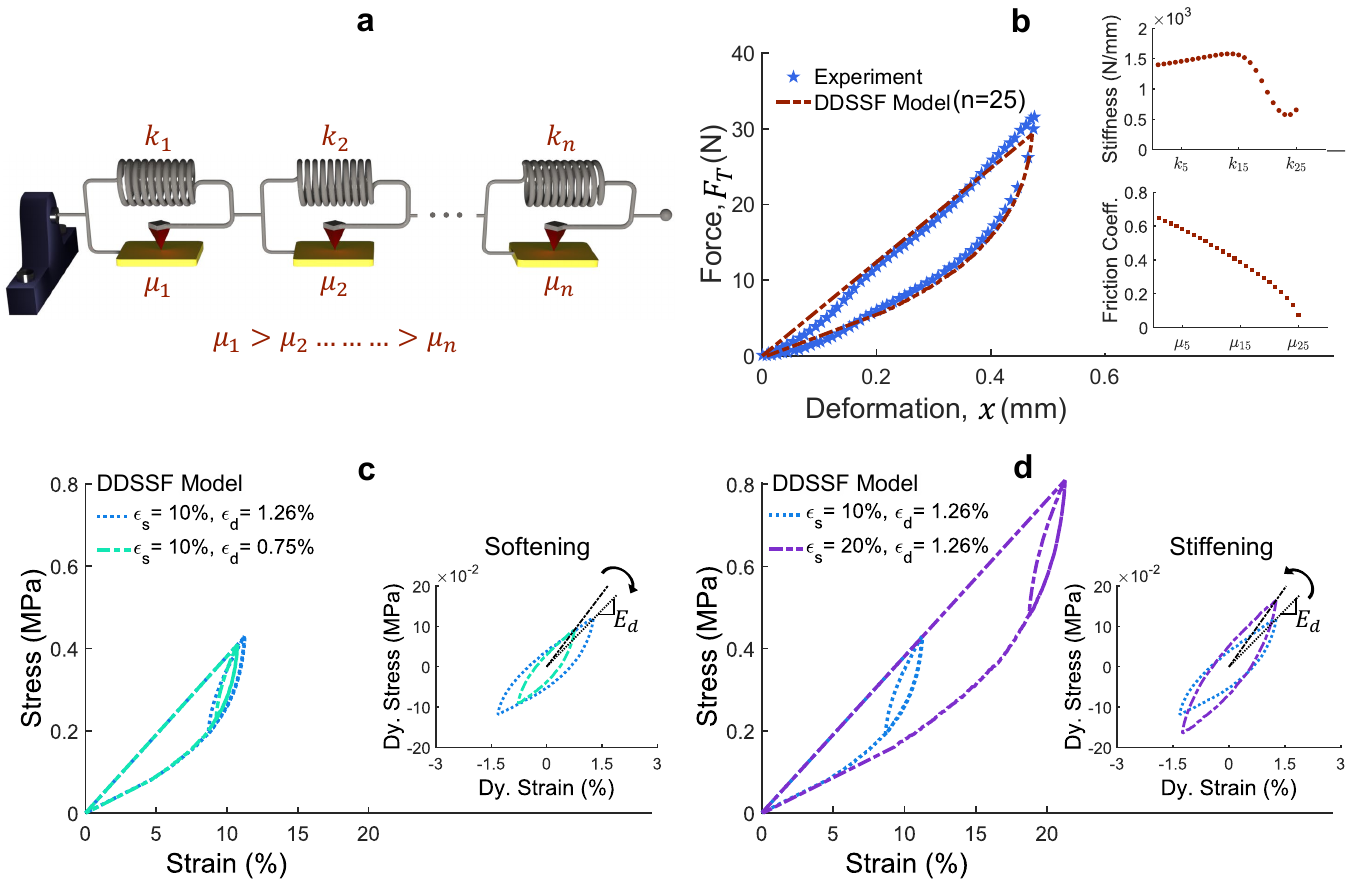}
	\caption{\textbf{DDSSF model validation.} (a) DDSSF model made of $n$ spring-slider pairs. (b) Force-deformation response of a $n=25$ DDSSF model, capturing the experimental hysteresis. Fitted springs stiffnesses and friction coefficients plotted in the inset. (c) Strain amplitude-dependent dynamic-softening and (d) static precompression-dependent stiffening captured by the DDSSF model.}
	\label{figs5}
\end{figure}

\Cref{fig4}(a) illustrates the experimental setup, featuring the VACNT–aluminum waveguide housed inside a bespoke PTFE (Polytetrafluoroethylene) sleeve. A screw–nut mechanism is mounted at one end to apply static precompression, while the other end features an impact anvil used to excite wave pulses. Using a gas gun setup (\Cref{figs6}(a)), we accelerated an aluminum striker to impact the anvil at various velocities $(v_i)$, generating traveling stress pulses with different amplitudes but similar frequency content (see Materials and Methods). We measured the evolution of the traveling pulse using two strain gauges mounted on aluminum cylinders separated by a known distance $\Delta L$ (see Materials and Methods). The effective pulse speed $(v_s)$, calculated by dividing $\Delta L$ by the time interval $\Delta t$ measured from the strain gauge signals (\Cref{figs6}(b)), is plotted in \Cref{fig4}(b) as a function of $v_i$ for various levels of the static precompressive strain applied to the layered system $(\epsilon_s)$. Increasing $\epsilon_s$ stiffens the VACNT samples, raising the baseline wave speed. In contrast, increasing $v_i$ subjects the VACNTs to higher dynamic strain $(\epsilon_d)$, causing softening and thereby decreasing the wave speed. The dependence of wave speed on both dynamic strain and static precompression follows trends similar to those observed in the dynamic modulus measurements (\Cref{fig2}(i)). To further validate these observations, we used the DDSSF model to estimate the group velocity, whose magnitude and trends closely match the experimental results (see details in \hyperref[section:sd2]{SI} and \Cref{figs10}(d)).

Notably, such unique wave speed tunability is not achievable in typical viscoelastic materials due to their characteristic hyperelastic nonlinearity, which tends to increase wave speed at higher strain amplitudes \cite{fan2012nonlinear}. Even nominally softening viscoelastic systems can exhibit apparent strain-rate dependent stiffening with increasing impact velocity, which can counteract any softening effects that may be present. Additionally, their inherent frequency dependence introduces additional dispersion and fading memory, making robust wave-speed regulation extremely challenging. In contrast, VACNT foams offer precise wave speed tunability under repeated excitations due to their robust rate independence, ability to sustain a precompressed stress-state without relaxation, and consistent return to this state enabled by RPM. When combined with dual dynamic modulus tunability, these properties unlock a unique class of metamaterials based on VACNT-enabled elastic waveguides and modulators, suited for various vibration control, shock mitigation, and analog computing applications \cite{mousa2024parallel}.

\subsection*{Engineering opportunities with VACNT-enabled elastic wave modulators}

The predictive control of elastic wave speed enables wavefront shaping by fine-tuning phase gradients in acoustic devices and other engineered wave systems \cite{guo2023development}. With embedded amplitude-sensitive layers, these gradients can be made to self-regulate based on the amplitude of the incoming pulse. This capability facilitates the design of wave limiters for protective systems that neutralize high-amplitude loads, as well as analog matched filters that identify target waveforms without the need for costly digital computation. To demonstrate this concept using VACNT-enabled periodic waveguides, we analytically modeled a phased array composed of individual VACNT–aluminum layered systems arranged in a radially symmetric pattern with uniform spacing, as illustrated in \Cref{fig4}(c). Static precompression---that increases with distance from the center---is applied respecting the radial symmetry of the phased array. As a result, an excitation pulse of a given amplitude incident uniformly across the entire phase array travels slowest at the center array and progressively faster in the outer arrays, thereby establishing a radial phase gradient across the array (\Cref{fig4}(c)).

In \Cref{fig4}(b), the individual curves corresponding to each precompression level are more widely spaced at low $v_i$ and converge as $v_i$ increases, indicating that the phase gradient is sensitive to incident wave amplitude. To demonstrate this, we applied a uniform incident pulse to each array from one end (\Cref{fig4}(c)), allowing the pulses to propagate through the array and transmit elastic waves into a receiver medium placed at the opposite end. Due to the higher precompression in the outermost arrays, those pulses arrive first, while the center array delivers its pulse last.
A low-amplitude incidence preserves a larger speed contrast between the arrays, causing the waves generated in the receiver medium to coalesce and form a sharp pressure peak at the focal point $\mathcal{F}$ (\Cref{fig4}(d,e)). In contrast, a high-amplitude incidence reduces the speed contrast, disrupting the wave focusing and resulting in an attenuated pressure output at the same location (\Cref{fig4}(f,g)).
This system functions as an amplitude-dependent wave limiter: only a pre-selected low incident amplitude results in constructive coalescence, while higher amplitudes lose synchrony, spread out over time, and emerge significantly attenuated. By tuning the chains to coalesce at this low amplitude (see details in \hyperref[section:sd2]{SI}), the system can suppress a broad range of higher-amplitude incident waves---enabling smart, protective shock-absorbing liners that redistribute hazardous impact energy both spatially and temporally above a specified threshold. This novel engineered material design exploiting VACNT's RPM should be understood in contrast to conventional wave limiters that operate in linear regime and the nonlinear systems that undergo plasticity and permanent damage under higher intensity loads \cite{yang2012amplitude,mork2022nonlinear}. The robust tunability of the proposed periodic waveguide and phased array system arises from the constitutive memory of VACNT foams: the return point memory (RPM) generated by deformation-dependent stick–slip friction, which resets each layer to its original stress state after every pulse---unlike viscoelastic polymers whose stresses gradually fade (memory loss).

Similar principle unlocks the potential for wave-based mechanical analog computing in matched filtering, where the goal is to maximize the signal-to-noise ratio by cross-correlating an incoming waveform with a time-reversed template \cite{zangeneh2021analogue}. The phased array can be configured to produce a sharp peak only when the incident pulse matches a target amplitude profile. A specific wave-speed distribution can be achieved by tailoring various design parameters, and with further adjustment of the receiver medium, a desired focal point can be realized (see details in \hyperref[section:sd2]{SI}). The result is a passive, mechanical matched filter that amplifies the desired waveform while flattening and attenuating all others \cite{mu1988amplitude}---importantly without the pitfalls of latency and cost of digital computations.

\section*{Discussion}

We demonstrated, for the first time, constitutive return point memory (RPM) in an elastically recoverable VACNT foam. In contrast to contemporary geometric bistability-based memory in metamaterials, RPM is the material's ability to remember deformation history in the stress-strain hysteresis---analogous to magnetic RPM. We developed a novel frictional damping model that reveals RPM originates from the memory of loading history encoded in nanoscale stick-slip frictional interactions. We showed that the interplay between RPM and frictional interactions impart dual tunability to the VACNT’s dynamic modulus—enabling stiffening via precompression and softening via dynamic strain.

We leverage the tunable dynamic modulus to experimentally demonstrate stress wave propagation in a periodic VACNT array waveguide with rigid interlayers, with wave speed independently adjustable by dynamic input amplitude and precompression---a capability uniquely enabled by the robust RPM effects in VACNTs. Arranging these individual waveguides in radial symmetry creates a VACNT-enabled phased array, where wave speed tunability is used to achieve phase contrast in wave propagation. This allows individual stress waves to focus at a certain input amplitude while diffusing and attenuating at higher amplitudes, enabling amplitude-dependent shock limiters.

Additionally, the fatigue resistance \cite{suhr2007fatigue} and thermal stability \cite{xu2010carbon,chawla2024superior} of VACNTs ensure reliable operation under repeated loading and in extreme environments. These attributes position VACNTs as an enabling material for next-generation adaptive shock absorbers, ultrasound imaging, vibration isolation, analog signal processing, and a wide range of real-time analog computing applications.

Due to limitations imposed by viscoelastic fading memory effects and plastic deformation, the unique properties of VACNTs cannot be replicated in traditional material systems. However, fundamental insights from the DDSSF model suggest that similar behavior can be achieved in nanostructured materials whose dissipative response is governed by nanoscale friction rather than viscoelasticity---including boron nitride nanotubes, fibrous biological materials, and nanoarchitected materials with engineered frictional dissipation.

\section*{Materials and Methods}
\subsection*{CVD synthesis}

We synthesize VACNT foams using a floating catalyst thermal chemical vapor deposition (tCVD) process, with toluene as the carbon source and ferrocene as the catalyst precursor. Our tCVD setup consists of a $1100\;mm$ long fused quartz glass tube (Technical Glass Products $50 \times 55$) enclosed with custom-made sanitary fittings attached at both ends. For synthesis, the glass tube is placed inside a tube furnace (Carbolite Gero CTF 12/75/700) maintained at a temperature of $1100\;K$. A $20 \times 20\;mm$ p-type silicon wafer substrate is positioned in the furnace’s peak heating zone (12 inches from the inlet) to facilitate CNT growth. Steel fittings are then attached at both ends, and argon gas is flowed at $760\;sccm$ from the inlet to create an inert atmosphere inside the tube at 1 atm pressure. The gas exits through a glass bubbler connected at the outlet.
Once the air inside the enclosure is fully replaced with argon, hydrogen gas is introduced at $40\;sccm$, making the carrier gas mixture 5\% hydrogen and 95\% argon. Hydrogen enhances catalyst activity by reducing amorphous carbon formation, resulting in purer CNTs. The flow of the carrier gas mixture is regulated using individual mass flow controllers (MKS G-series GE50 A). To initiate the synthesis, a feedstock solution of $0.8\;g$ ferrocene in $80\;ml$ toluene ($[w/v] = 0.01\;g/ml$) is supplied at $0.8\; ml/min$ using a glass syringe and syringe pump (NE-1000) connected to the inlet. The CNTs grow vertically aligned on the substrate, reaching a height of approximately $1.5-3\;mm$.
Once the feedstock solution is depleted, the furnace is switched off, and the hydrogen flow is stopped while argon continues to flow, allowing the furnace to cool down to approximately $700\;K$. At this temperature, the substrate with a $1-3\;mm$ thick VACNT film is retrieved and prepared for mechanical characterization. The film is separated and cut into smaller pieces: a biopsy punch is used to cut circular samples, and a microtome blade is used for square samples. The density of the samples typically ranges between $0.13\;g/cm^3$ and $0.26\;g/cm^3$.

\subsection*{Uniaxial quasistatic compression}

We conducted quasistatic compression and stress relaxation experiments using a commercial load frame, an Instron Electropulse E3000, equipped with a 250 N load cell. From a CVD-synthesized VACNT film measuring $\sim 20 \times 20\; mm$ and $1.59\;mm$ in thickness, we cut a square cross-section sample with dimensions of $5.09 \times 5.08\;mm$. First, we preconditioned the pristine sample by applying cyclic ramp loading at a strain rate of $0.01\; {s}^{-1}$ for 5 cycles, recording both the loading and unloading responses. On the preconditioned sample, exhibiting stable stress-strain behavior, we performed cyclic quasistatic compression at various strain rates.

\subsection*{Stress-relaxation experiments}

For consistent comparison, we performed the stress relaxation experiments on the same sample used for quasistatic characterization. To measure stress relaxation for various values of compressive strain, we applied ramp strain loading at a strain rate of $0.05 \; {s}^{-1}$, which was chosen because it is slow enough to reduce overshoot in strain due to the load frame's inertia but fast enough to quickly reach the desired strain level. We applied the strain up to the desired compression strain value and held it constant for 30 minutes. Since the majority of stress relaxation in viscoelastic materials occurs within the first 20 to 30 minutes, a duration of 30 minutes is sufficient to ascertain the occurrence of any relaxation. Additionally, since we found that VACNT foams do not exhibit any relaxation, extending the relaxation period will not provide any new information.

\subsection*{Dynamic mechanical analysis (DMA)}

We conducted broadband DMA experiments using a custom-built dynamic mechanical analyzer, consisting of a piezoelectric actuator (Physik Instrumente P841.10) and a dynamic force sensor (PCB Piezotronics 208C01). The actuator's motion is controlled by a position servo control module and an amplifier (Physik Instrumente E509 and E505, respectively). The actuator is attached to a fixed steel block, while the force sensor is mounted on a linearly displaceable steel block facing the actuator.
To perform DMA measurements, the preconditioned sample is first attached to the actuator tip and then precompressed to a desired strain between the actuator and the force sensor by adjusting the force sensor’s block using a micrometer head. Once the sample is precompressed, we apply 20 cycles of sinusoidal strain with the desired amplitude and frequency by supplying a voltage waveform to the actuator using an analog output device (National Instruments, NI-9269). The force sensor records the force output via a signal conditioner (PCB 482C15) and an analog input device (NI 9215).
After collecting the 20-cycle signal, the first 5 and last 5 cycles are removed, and the average of the middle 10 cycles is taken to obtain the hysteresis response. From this hysteresis response, the dynamic modulus and loss factor are measured through amplitude and frequency sweeps. 

While our custom DMA setup enables high-frequency measurements up to $f = 1\;kHz$, the available amplitude range decreases with frequency, being limited to $7.6\;\mu m$ for $f < 100\; Hz$ and reducing to $2\;\mu m$ as the frequency approaches $1\; kHz$. However, the frequency-independent behavior of VACNTs allows us to measure the dynamic modulus and loss factor as functions of amplitude at low frequencies, with the results applicable to all frequencies. Thus, we conducted amplitude sweep experiments at $f = 10\;Hz$ for amplitudes up to $7.6\;\mu m$. For further higher amplitudes, up to $20\;\mu m$, we used the Instron E3000 load frame, as described above, at a frequency of $0.02\;Hz$, leveraging the frequency-independent behavior.

\subsection*{Wave propagation experiment}

\begin{figure}[H]
	\centering
	\includegraphics[width=\textwidth]{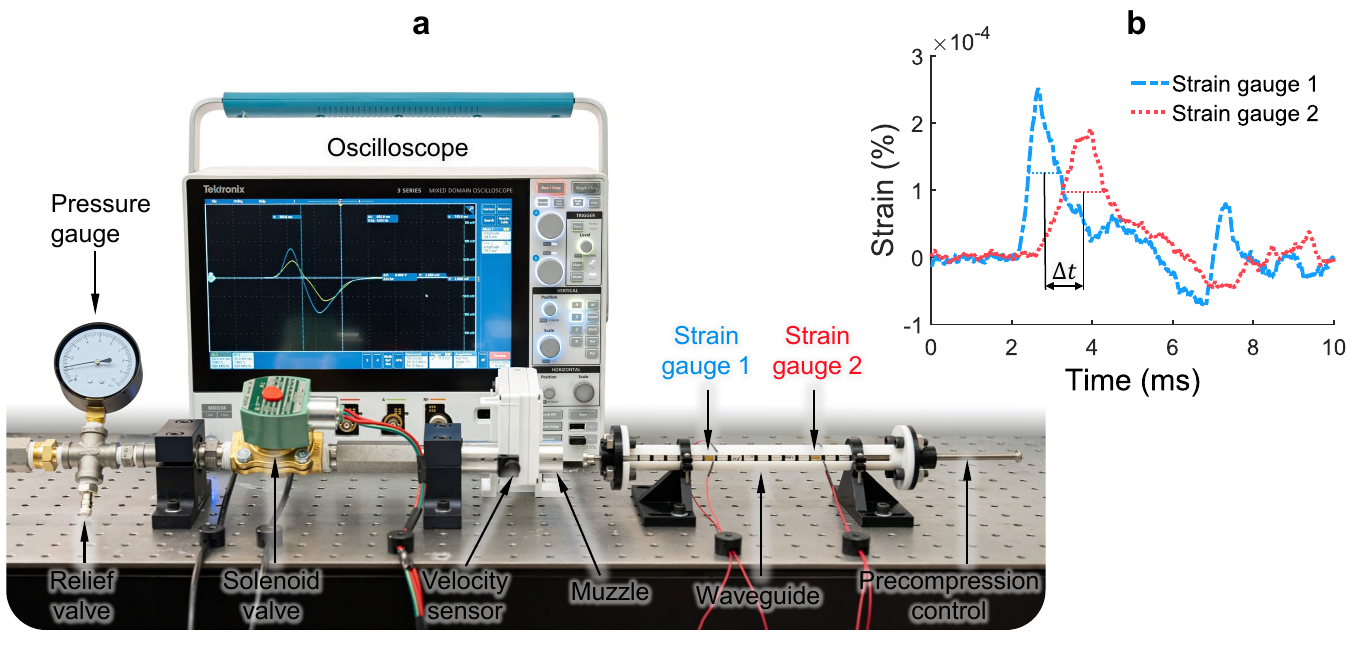}
	\caption{\textbf{Wave propagation experiment.} (a) Picture of the experimental setup used to measure wave propagation in VACNT-aluminum periodic waveguide. (b) Representative strain signals measured as function of time from two strain gauges}
	\label{figs6}
\end{figure}

We constructed the periodic waveguide using 17 VACNT foam samples, each with an average thickness of $2.06\;mm$ and a diameter of $5\;mm$, alternately layered with $17$ aluminum cylinders measuring $8\;mm$ in length and $6\;mm$ in diameter (\Cref{figs6}(a)). We selected VACNT samples with similar densities and stress–strain responses to ensure uniform precompression across all layers. Before assembling the waveguide, we preconditioned each foam sample up to $50\%$ strain. We used an aluminum cylinder of identical dimensions as a projectile, launching it with a custom-built gas gun powered by an electric air compressor. The projectile’s design produced a wavelength comparable to the structural periodicity, placing wave propagation in the short-wavelength regime where dispersion dominates. By adjusting the compressor pressure, we controlled the projectile’s velocity. Notably, increasing the amplitude of the incident stress pulse did not significantly alter its frequency content, allowing us to isolate amplitude effects on wave propagation for signals with similar spectral characteristics.

To capture the time evolution of the traveling stress pulse, we mounted strain gauges (Kyowa KSPB-1-350-E4) on the $6^{th}$ and $14^{th}$ aluminum interlayers. These gauges were configured in a quarter-bridge circuit and balanced using individual signal conditioning amplifiers (Vishay 2310B). Signals were recorded using an NI cDAQ-9174 chassis and an analog input module (NI 9215), operated through MATLAB’s Data Acquisition Toolbox. We measured the projectile speed $(v_i)$ just before impact using a PASCO wireless smart gate (PS-3225). To calculate the effective wave speed ($v_s$), we divided the known separation distance between the strain gauges---measured after applying precompression---by the time delay (\Cref{figs6}(b)) between the strain gauge signals---measured using the FWHM middle-point method. For each precompression level, we repeated the experiment for every input velocity three times to test repeatability. 

\bibliographystyle{unsrtnat}   
\bibliography{citations.bib}

\newpage

\section*{Acknowledgement}

This research is supported by the U.S. Office of Naval Research under PANTHER program (Award number N000142412200) through Dr. Timothy Bentley as well as by the solid mechanics program of the U.S. Army Research Office (Award number W911NF2010160) through Dr. Denise Ford.

\section*{Author Contributions}

A.G. conducted the quasistatic compression, stress relaxation, and DMA experiments on VACNT foams. A.G. and B.M. carried out the wave propagation experiments and developed the frictional damping model. B.M. performed the analysis for phased array system and wave coalescence. N.J. conducted preliminary quasistatic compression and stress-relaxation experiments. K.C. synthesized the VACNT samples. A.G., B.M., and R.T. analyzed the results and drafted the manuscript, with input from all authors. A.G. and B.M. contributed equally.

\section*{Competing Interests}
The authors declare no competing interest.

\section*{Data Availability}
The raw data generated in this study is available in a public GitHub repository (\href{https://github.com/ThevamaranLab/DDSSF-Model.git}{https://github.com/ThevamaranLab/DDSSF-Model.git})

\label{section:data}

\section*{Code Availability}
Our MATLAB scripts for fitting and evaluating the DDSSF model are available in a public GitHub repository (\href{https://github.com/ThevamaranLab/DDSSF-Model.git}{https://github.com/ThevamaranLab/DDSSF-Model.git})

\label{section:sd4}

\newpage

\section*{Supplementary Text}

\label{section:sd2}

\subsection*{Discrete form of Boltzmann superposition integral}

The time-dependent stress-relaxation function of a standard linear solid (SLS) is given by the following expression \cite{lakes2009viscoelastic}:

\begin{equation}
    E\left(\bar{t}\right)=E_{\infty}+E_t\times e^{-\bar{t}}
\end{equation}

Here, $\bar{t} = t/t_r$ denotes a dimensionless time, where $t_r$ is the relaxation time. Any arbitrary strain applied as a function of time, $\epsilon(\bar{t})$, can be expressed as a series of discrete strain increments, $d\epsilon_i$, applied at specific times, $\bar{\tau}_i$.

\begin{equation}
    \epsilon\left(\bar{t}\right) = \sum_{\bar{\tau}_i=0}^{{\bar{\tau}_i=\bar{t}}} d\epsilon_i\times\mathcal{H}\left(\bar{t}-\bar{\tau}_i\right)
\end{equation}

Where $\mathcal{H}$ denotes the Heaviside step function. For a strain increment of $d\epsilon_i$, the corresponding stress increment can be expressed in dimensionless form as follows

\begin{equation}
    {d\bar{\sigma}_i}(\bar{t})= d\epsilon_i\times\mathcal{H}\left(\bar{t}-\bar{\tau}_i\right)\times \left( 1+\frac{E_t}{E_\infty} \times e^{-\left(\bar{t}-\bar{\tau_i}\right)} \right) \;\;\;,\;\;\bar{\sigma}_i=\frac{{\sigma}_i}{E_{\infty}}
\end{equation}
\begin{equation}
   \boxed{{d\bar{\sigma}_i}(\bar{t})= d\epsilon_i\times \mathcal{W}\left(\bar{t}-\bar{\tau}_i \right)}
\end{equation}

Where $\mathcal{W}\left(\bar{t} - \bar{\tau}_i\right)$ represents a weight associated with $d\epsilon_i$. The weight decreases with time, leading to the fading memory effect. The total stress as a function of time can be obtained by adding all the stress increments

\begin{equation}
    \bar{\sigma}(\bar{t}) = \sum d\bar{\sigma}_i \left(\bar{t} \right)
\end{equation}

\subsection*{Absence of stress-relaxation behavior}

\begin{figure}[H]
	\centering
	\includegraphics[width=\textwidth]{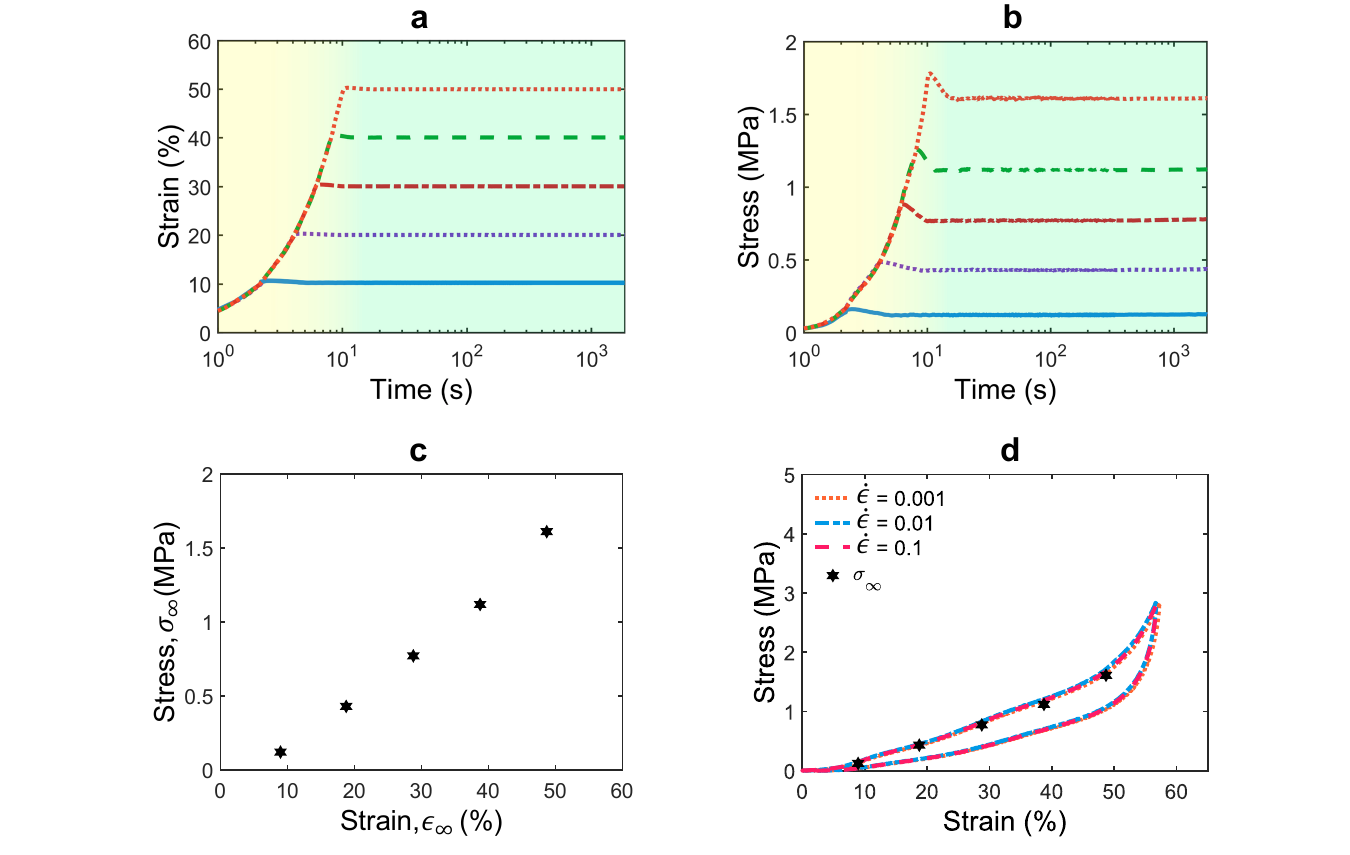}
	\caption{\textbf{Stress relaxation experiments.} (a) Strain applied over time up to various equilibrium strain levels and the corresponding (b) stress responses. (c) Equilibrium stress $(\sigma_{\infty})$ plotted against the corresponding equilibrium strain $(\epsilon_{\infty})$. (d) Equilibrium stress and strain values overlaid on the preconditioned quasistatic stress–strain curve.}
	\label{figs7}
\end{figure}

Under quasistatic compression, VACNT foams demonstrated a clear rate independent behavior (see Figure 2(d) in the manuscript). To further validate their strain-rate-independent behavior, we performed stress-relaxation experiments on these VACNT samples by swiftly ramping the strain to a desired level $(\epsilon_{\infty})$ and holding it constant for 1800 seconds (30 minutes) (\Cref{figs7}). \Cref{figs7}(b) shows the corresponding stress response as a function of time for various constantly held strains. Ignoring the slight overshoot in stress during ramping, caused by the inertia of the moving crosshead, the stress remains constant for the duration the strain is held constant (green shaded region in \Cref{figs7}(a,b)), suggesting no stress-relaxation behavior. In \Cref{figs7}(c), we plot this constant stress (or equilibrium stress, $\sigma_{\infty}$) as a function of constantly held strain ($\epsilon_{\infty}$), demonstrating an approximately linear relationship. Overlaying on the quasistatic stress-strain hysteretic response (\Cref{figs7}(d)), the $\sigma_{\infty}$ and $\epsilon_{\infty}$ data coincide on the loading curve. This suggests that during the ramping phase of strain in the relaxation experiment, the stress increases, tracing the quasistatic loading curve. Once the strain is brought to a hold, the stress also becomes constant at that point and remains unchanged. Conversely, if a load-controlled experiment were performed, where stress is held constant (\Cref{figs7}(b)) and strain is monitored, \Cref{figs7}(a) suggests that the strain would remain constant, indicating no creep. The absence of stress relaxation and creep in VACNT foams results in a persistent memory effect, as the influence of an applied strain does not diminish over time---unlike fading memory effect in viscoelastic foams.

\subsection*{Scaling of the quasistatic unloading curve}

\begin{figure}[H]
	\centering
	\includegraphics[width=\textwidth]{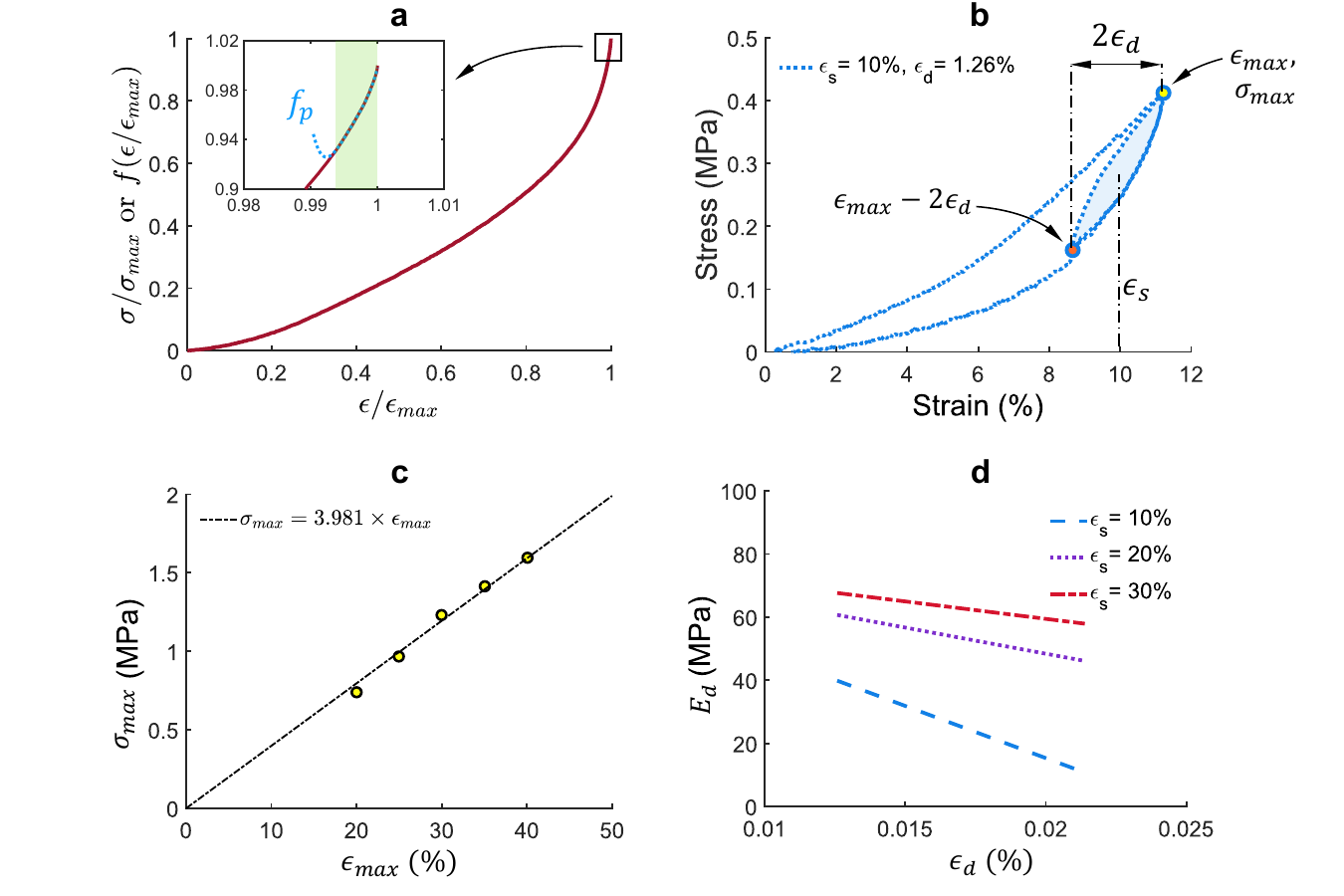}
	\caption{ \textbf{Unloading curve scaling leads to softening and stiffening.} (a) Normalized global hysteretic unloading curve with a polynomial fit near the maximum strain. (b) Global hysteretic loop and subloop under dynamic loading. (c) Maximum stress versus maximum strain with linear fit. (d) Approximate dynamic modulus as a function of dynamic strain amplitude and quasistatic compression strain. }
	\label{figs8}
\end{figure}

In the quasistatic cyclic-ramp compression experiments, the unloading curves measured after loading the VACNT sample to different maximum strains $(\epsilon_{max})$ were found to scale with the magnitude of $\epsilon_{max}$ while preserving their overall shape. Consequently, as shown in \Cref{figs2}(c), the unloading curves for different $\epsilon_{max}$ nearly overlap when the stress and strain are normalized by their respective maximum values. In \Cref{figs8}(a), we present the average of the normalized unloading curves shown in \Cref{figs2}(c). Assuming that the normalized unloading curve is a continuous function of normalized strain $(\epsilon/\epsilon_{max})$, we have $\sigma/\sigma_{max} = f(\epsilon/\epsilon_{max})$. This representation enables us to express the general form of the unloading curve as follows:

\begin{equation}
    \sigma = \sigma_{max}\times f\left( \frac{\epsilon}{\epsilon_{max}} \right)
    \label{eqscals}
\end{equation}

where $f(0)=0$, $f(1)=1$, and $\sigma_{max}$ is the maximum stress corresponding to $\epsilon_{max}$ \\

In a DMA experiment involving the application of a static precompression strain $\epsilon_s$ and a dynamic sinusoidal strain of amplitude $\epsilon_d$ (\Cref{figs8}(b)), the maximum strain is given by 

\begin{equation}
    \epsilon_{max}=\epsilon_s+\epsilon_d
    \label{epmax}
\end{equation}

The dynamic modulus $(E_d)$, is defined as the ratio of the total change in stress to the total change in strain within the dynamic hysteresis loop (\Cref{figs8}(b)), and is expressed as follows:

\begin{equation}
    E_d=  \frac{\sigma_{max}-\sigma_{min}}{2\epsilon_d} \;\;\;\;,\;\;\;\sigma=\left\{\begin{array}{l}\sigma_{\max } \;\;\textnormal{at}\;\; \epsilon=\epsilon_{\max } \\ \sigma_{\min }\;\;\textnormal{at}\;\;\epsilon=\epsilon_{\max }-2 \epsilon_d\end{array}\right.
    \label{Ed}
\end{equation}

Substituting (\ref{eqscals}) in (\ref{Ed})

\begin{equation}
    E_d=  \frac{1}{2\epsilon_d} \times \left[\sigma_{max} - \sigma_{max}\times f \left( \frac{\epsilon_{max}-2\epsilon_d}{\epsilon_{max}} \right) \right]
    \label{Ed1}
\end{equation}

\begin{equation}
    E_d=  \frac{1}{2\epsilon_d} \times \left[\sigma_{max} - \sigma_{max}\times f \left( \frac{\epsilon_s-\epsilon_d}{\epsilon_s+\epsilon_d} \right) \right]
    \label{Ed2}
\end{equation}

\begin{equation}
    E_d=  \frac{\sigma_{max}}{2\epsilon_d} \times \left[ 1-  f \left( 1-\frac{2\epsilon_d}{\epsilon_s+\epsilon_d} \right) \right]
    \label{Ed3}
\end{equation}

Expanding the perturbed function term in the equation above using the Taylor series yields:

\begin{equation}
    f \left( 1-\frac{2\epsilon_d}{\epsilon_s+\epsilon_d} \right) = f(1)-\frac{2\epsilon_d}{\epsilon_s+\epsilon_d}f'(1)+\frac{1}{2} \left( \frac{2\epsilon_d}{\epsilon_s+\epsilon_d}\right)^2 f''(1) ............
    \label{Ed4}
\end{equation}

Substituting $f(1)=1$ and neglecting higher-order terms for small dynamic strain amplitudes, $\frac{2\epsilon_d}{\epsilon_s+\epsilon_d}\ll1$:

\begin{equation}
    f \left( 1-\frac{2\epsilon_d}{\epsilon_s+\epsilon_d} \right) \approx 1-\frac{2\epsilon_d}{\epsilon_s+\epsilon_d}f'(1)+\frac{2\epsilon_d^2}{(\epsilon_s+\epsilon_d)^2} f''(1)
    \label{Ed5}
\end{equation}

Substituting above expression in (\ref{Ed3})

\begin{equation}
    E_d\approx  \frac{\sigma_{max}}{2\epsilon_d} \times \left[\frac{2\epsilon_d}{\epsilon_s+\epsilon_d}f'(1)-\frac{2\epsilon_d^2}{(\epsilon_s+\epsilon_d)^2} f''(1) \right]
    \label{Ed6}
\end{equation}

\begin{equation}
    E_d\approx  \frac{\sigma_{max}}{(\epsilon_s+\epsilon_d)} \times \left[f'(1)-\frac{\epsilon_d}{(\epsilon_s+\epsilon_d)} f''(1) \right]
    \label{Ed7}
\end{equation}

\begin{equation}
    E_d\approx  \frac{\sigma_{max}}{\epsilon_{max}} \times \left[f'(1)-\frac{\epsilon_d}{(\epsilon_s+\epsilon_d)} f''(1) \right]
    \label{Ed8}
\end{equation}

For maximum strain below the onset of densification (i.e., $\epsilon_{max} <48.42\%$), the loading curve is nearly linear (see \Cref{figs2}(b)), resulting in $\sigma_{max}$ being linearly proportional to $\epsilon_{max}$. \Cref{figs8}(c) illustrates this linear relationship, where $\sigma_{max} = 3.981 \times \epsilon_{max}$ for $20\% \leq \epsilon_{max} \leq 40\%$. For $\epsilon_{max}<20\%$, the loading curve is slightly nonlinear due to the uneven surface of the sample

\begin{equation}
    E_d\approx  3.981 \times \left[f'(1)-\frac{\epsilon_d}{(\epsilon_s+\epsilon_d)} f''(1) \right]
    \label{Ed9s}
\end{equation}

From a visual inspection of the normalized average unloading curve (\Cref{figs8}(a)), it is evident that $f'(1) > 0$ and $f''(1) > 0$. In (\ref{Ed9s}), it is clear that for positive values of $f'(1)$ and $f''(1)$, $E_d$ will increase with $\epsilon_s$, while $\epsilon_d$ will decrease, which is consistent with our experimental observations. To estimate the values of $f'(1)$ and $f''(1)$, we fit a polynomial to approximate the initial portion of the unloading curve, as shown in the inset of \Cref{figs8}(a). The range of $\epsilon/\epsilon_{max}$ where the polynomial exactly overlaps the curve is shaded in green. The following is the approximate polynomial fit:

\begin{equation}
    f_p\left( \frac{\epsilon}{\epsilon_{max}}\right)=10^7\times\left(4.28\left( \frac{\epsilon}{\epsilon_{max}}\right)^4-17.07\left( \frac{\epsilon}{\epsilon_{max}}\right)^3+25.5\left( \frac{\epsilon}{\epsilon_{max}}\right)^2-16.93\left( \frac{\epsilon}{\epsilon_{max}}\right)+4.22 \right)
    \label{fp}
\end{equation}

For small dynamic strain amplitudes, the normalized unloading curve can be approximated using the polynomial described above.

\begin{equation}
    f\left( \frac{\epsilon}{\epsilon_{max}}\right) \approx f_p\left( \frac{\epsilon}{\epsilon_{max}}\right)  \;\;,\;\;\epsilon_d\ll\epsilon_s
\end{equation}

By taking the derivative of the polynomial and substituting $\epsilon/\epsilon_{max} = 1$, we can obtain approximate values for $f'(1)$ and $f''(1)$. This allows us to rewrite (\ref{Ed9s}) as follows:

\begin{equation}
    E_d\approx  3.981 \times \left[20.49-\frac{\epsilon_d}{(\epsilon_s+\epsilon_d)} 8332.95 \right]
    \label{Ed10}
\end{equation}

In \Cref{figs8}(d), we plot $E_d$ as a function of $\epsilon_s$ and $\epsilon_d$ calculated using the above approximate relation. Clearly, $E_d$ increases with $\epsilon_s$, indicating dynamic stiffening, while it decreases with $\epsilon_d$, indicating dynamic softening. Although these results are valid only for $\epsilon_d \ll \epsilon_s$, the magnitudes of $E_d$ are consistent with direct experimental measurements (Figure 2(i) in the manuscript). While including more terms in the Taylor series expansion could improve (\ref{Ed10}) for larger $\epsilon_d$, a much higher-order polynomial fit would be required to cover a larger range of $\epsilon/\epsilon_{max}$ as shown in \Cref{figs8}(a) (inset), which is impractical. On the other hand, the experimentally measured unloading curve is not smooth enough to allow for higher-order numerical derivatives. Nevertheless, the analysis in this section reveals the relationship between the unique scaling property of the unloading curve observed during quasistatic ramp compression and the dynamic amplitude-dependent softening, as well as the static precompression-dependent stiffening observed in DMA measurements of the preconditioned VACNT foam.

\subsection*{DDSSF model formulation}

Consider $n$ springs with stiffnesses $k_i$ ($i = 1, 2, \dots, n$) and $n$ Coulomb sliders with friction coefficients $\mu_i$ ($i = 1, 2, \dots, n$). Each spring ($k_i$) is connected in parallel with its corresponding Coulomb slider ($\mu_i$), forming $n$ spring-slider pairs, which are then connected in series. Assuming the deformation in the $i^{th}$ spring-slider pair is $x_i$, the resultant force in the spring $(F_{s,i})$, the frictional force in the corresponding slider $(F_{bf,i})$, and the total force $(F_T)$ are given as follows:

\begin{equation}
    F_{s,i}=k_i x_i
\end{equation}
\begin{equation}
    F_{bf,i}=-\mu_i \times F_{s,i} \times \textnormal{sign}(\dot{x}_i)
\end{equation}
\begin{equation}
    F_T=F_{s,i}+F_{bf,i}
\end{equation}

Here, $\textnormal{sign}(\;)$ denotes the signum function. To capture the increase in internal friction between nanotubes as they become more compacted under compression, the magnitude of the friction force in each slider is assumed to be proportional to the corresponding spring force ($|F_{bf,i}|\propto|F_{s,i}|$). Since $F_{s,i} \propto x_i$, the friction force in each slider becomes deformation-dependent. The negative sign in the expression for $F_{bf,i}$ indicates that the friction force always opposes the direction of deformation (sign($\dot{x}_i$)). During compression or loading ($\textnormal{sign}(\dot{x}_i) = -1$), the friction force acts in the same direction as the spring force (additive), whereas during decompression or unloading ($\textnormal{sign}(\dot{x}_i) = +1$), it acts in the opposite direction (subtractive), as expressed below

\begin{equation}
    F_T=k_i x_i (1+\mu_i)\;\;(\textnormal{Loading})  ,\;\;F_T=k_i x_i (1-\mu_i)\;\;(\textnormal{Unloading})
\end{equation}

Due to the series configuration, the total force $F_T$ is the same across all spring-slider pairs. When an external compressive force is applied and increased monotonically starting from $F_T = 0$, the break-free friction in each slider---initially zero---offers no resistance to deformation. As a result, all spring-slider pairs begin to compress immediately upon the application of $F_T$. At an instance, the deformation in each pair $(x_{i})$ as well as the total deformation of the model ($x$), are given as follows

\begin{equation}
    x_{i}= \frac{F_{T}}{k_i(1+\mu_i)}
\end{equation}
\begin{equation}
    x= \sum_{i=1}^{n} x_i =F_{T} \sum_{i=1}^{n} \frac{1}{k_i(1+\mu_i)}
\end{equation}

Using the two equations above, the effective stiffness of the loading curve (\Cref{figs9}(a)) in the model’s global hysteresis loop ($k_L$) can be expressed as follows

\begin{equation}
    \frac{1}{k_L}=\frac{x}{F_{T}}=\sum_{i=1}^{n} \frac{1}{k_i(1+\mu_i)}
\end{equation}

When the model is loaded to a maximum force of $F_T = F_{T,{max}}$, the corresponding maximum deformation in each pair, the maximum force in each spring, and the maximum friction force in each slider are given as follows

\begin{equation}
    x_{i,max} = \frac{F_{T,max}}{k_i(1+\mu_i)}
\end{equation}
\begin{equation}
    F_{s,i,max} = k_ix_{i,max}
\end{equation}
\begin{equation}
    F_{bf,i,max}=\mu_i k _i x_{i,max} = F_{T,max}\times \frac{\mu_i}{(1+\mu_i)}
    \label{fricbf}
\end{equation}

Unlike loading, which begins at $F_T = 0$ with all sliders initially unlocked due to zero friction, unloading starts with all sliders initially locked because of nonzero friction. As the external load ($F_T$) is reduced during unloading, the sliders begin to unlock sequentially (\Cref{figs9}(a)) in the order of increasing $\mu_i$ ($\mu_n < \mu_{n-1} < \dots < \mu_1$). For an $i^{th}$ slider to unlock and begin decompressing, the direction of the friction force must reverse---from $+F_{bf,i,{max}}$ to $-F_{bf,i,{max}}$---which requires the external force to decrease by $2F_{bf,i,{max}}$.
Since the friction force magnitude $F_{bf,i,{max}}$ is smallest for the slider with the lowest $\mu_i$ (\Cref{fricbf}), the $n^{{th}}$ slider---having the lowest friction coefficient---unlocks when the external force drops to $F_{T,{max}} - 2F_{bf,n,{max}}$ (\Cref{figs9}(a)), allowing both the slider and its parallel spring to decompress.
As $F_T$ continues to decrease, the spring-slider pairs unlock sequentially, resulting in a progressively decreasing slope of the unloading curve (\Cref{figs9}(a)). Once all pairs are unlocked, the curve reaches a constant slope (\Cref{figs9}(a)) that persists until the force is fully unloaded to $F_T = 0$. This slope is given as follows
\begin{equation}
    \frac{1}{k_{UL}}=\sum_{i=1}^{n} \frac{1}{k_i(1-\mu_i)}
\end{equation}

Similarly, each time the loading direction is reversed, all sliders become locked and then unlock sequentially as the magnitude of the external force changes by twice the current friction forces in the sliders. 

Notably, \Cref{fricbf} indicates that the friction force in each slider $(F_{bf,i,{max}})$ at the onset of unloading is proportional to the maximum external force reached during loading, $F_{T,{max}}$. As a result, for higher $F_{T,{max}}$, each slider must overcome a proportionally larger friction force $(2F_{bf,i,max})$ to unlock during unloading (\Cref{figs9}(a)). Consequently, the unloading curve exhibits a scaling behavior, as it scales with $F_{T,{max}}$ while preserving its overall shape, as observed in experiments.

\subsubsection*{Calculating $k_i$ and $\mu_i$ from experimental hysteresis curve}

\begin{figure}[H]
	\centering
	\includegraphics[width=\textwidth]{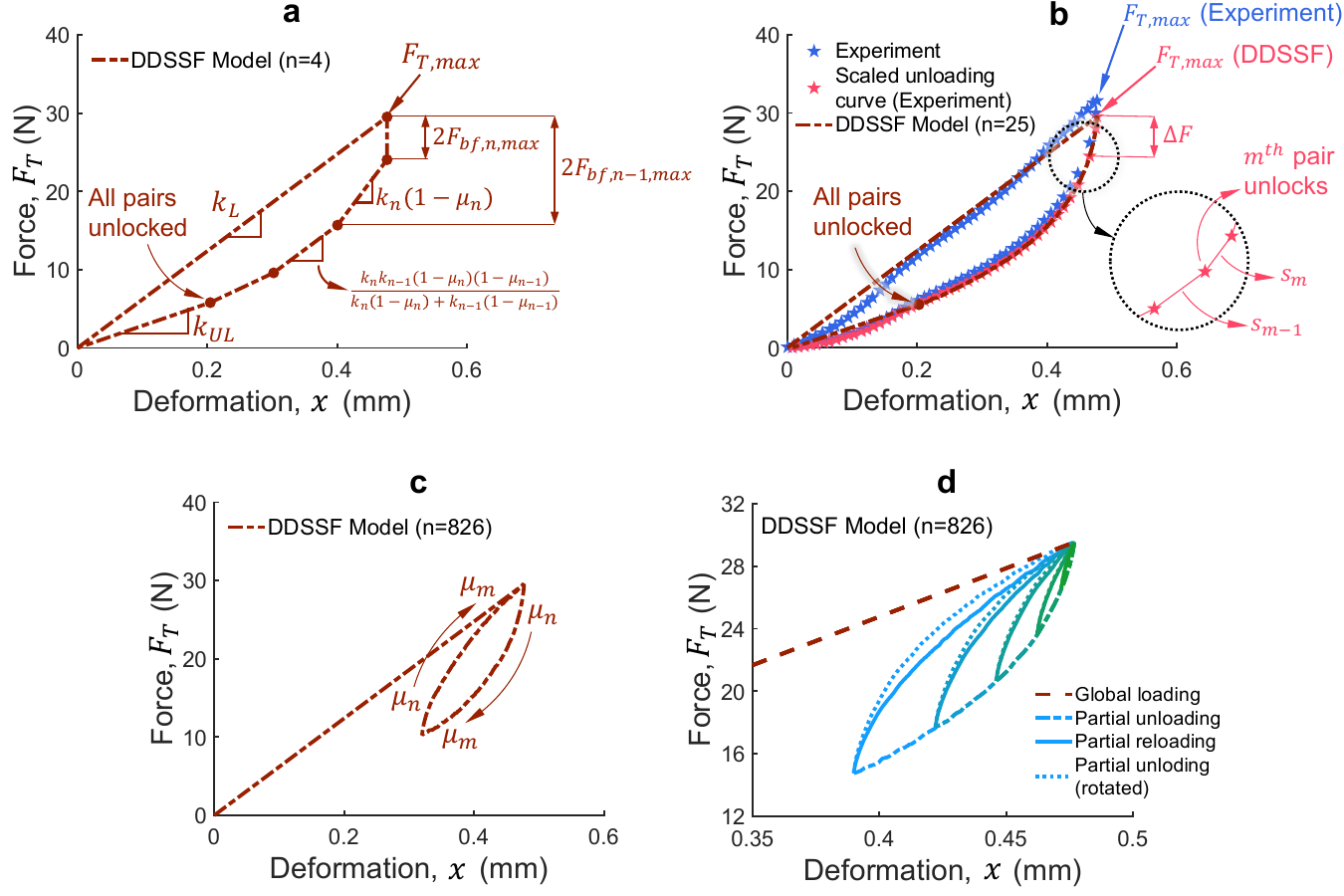}
	\caption{ \textbf{DDSSF model.} (a) Cyclic force-deformation response of the DDSSF model composed of four spring-slider pairs. (b) Experimental force-deformation response with the DDSSF model fitting process shown. (c) DDSSF model with $n = 826$ capturing return point memory. (d) Masing behavior ceases to exist for larger hysteretic sub-loops.}
	\label{figs9}
\end{figure}

For a given experimental cyclic global-hysteretic force–deformation response, the $k_i$ and $\mu_i$ values for the DDSSF model can be determined by dividing the experimental unloading curve into discrete segments. A larger number of segments results in more spring-slider pairs, leading to a smoother DDSSF model fit. \Cref{figs9}(b) shows the DDSSF model $(n=25)$ fitted to the experimental response obtained by compressing a VACNT sample to 30\% strain. The loading curve is first approximated using a linear fit with slope $k_L$. The maximum deformation applied experimentally is $0.477\;mm$, corresponding to 30\% strain on a $1.59\;mm$ thick sample. The value of $F_{T,max}$ used in the DDSSF model fit is calculated by multiplying $k_L$ with $0.477\;mm$. The experimental unloading curve is then scaled so that its maximum force matches the maximum force of the fitted loading curve (\Cref{figs9}(b)). 

As described earlier, during unloading, the spring-slider pairs unlock sequentially. On the scaled unloading curve, consider an instance when the total unloaded force is $\Delta F$ (\Cref{figs9}(b)), and the $m^{{th}}$ spring-slider pair unlocks, i.e., $2F_{bf,m,{max}} = \Delta F$ $(1<m<n)$. Using \Cref{fricbf}, the expression for $\mu_m$ can be written as follows

\begin{equation}
    \mu_m = \frac{\Delta F}{2F_{T,max}-\Delta F}
    \label{mum}
\end{equation}

As the $m^{{th}}$ spring-slider pair unlocks, the slope of the unloading curve changes (see inset of \Cref{figs9}(b)). Let the slope in the previous segment be $s_m$ and in the next segment be $s_{m-1}$. Their values can be expressed by the following equations

\begin{equation}
    \frac{1}{s_m} = \sum_{i=1}^{n-m} \frac{1}{k_{n+1-i}(1-\mu_{n+1-i})}
    \label{sm}
\end{equation}
\begin{equation}
    \frac{1}{s_{m-1}} = \sum_{i=1}^{n-m+1} \frac{1}{k_{n+1-i}(1-\mu_{n+1-i})}
    \label{sm-1}
\end{equation}

Subtracting \Cref{sm} from \Cref{sm-1} gives

\begin{equation}
    \frac{1}{s_{m-1}}-\frac{1}{s_{m}} = \frac{1}{k_m(1-\mu_m)}
\end{equation}

\begin{equation}
    k_m = \frac{s_{m-1}s_m}{(s_m-s_{m-1})\times(1-\mu_m)}
    \label{km}
\end{equation}

By substituting the value of $\mu_m$ from \Cref{mum} into \Cref{km}, the stiffness of the $m^{{th}}$ spring can be calculated. Similarly, by measuring the difference between the compliance (1/slope) of all other consecutive discrete segments of the scaled unloading curve, the stiffness values of all the springs $(n)$ can be determined. Whereas, the associated $\mu$ values for the Coulomb sliders can be obtained using \Cref{mum}, by inputting $\Delta F$ at various points along the scaled unloading curve. In \Cref{figs9}(b), the scaled unloading curve was fitted with the DDSSF model starting from the maximum point $x_{{max}}$ up to $x = 0.2\;{mm}$, beyond which the concavity of the curve changes due to nonlinearity in the loading curve (for $x < 0.2\;{mm}$). Nevertheless, if the loading curve was linear, the DDSSF model could be fitted over the entire unloading curve $(0 < x < x_{{max}})$.

Once the fitted values are known, the force–deformation response can be predicted for any input force–time waveform by tracking the magnitude and direction of the friction force in each slider. We implemented this algorithm in a MATLAB script that takes the fitted values of $k_i$ and $\mu_i$ as input to compute the force–deformation response for any applied force–time waveform (\hyperref[section:sd4]{GitHub}). As shown in the manuscript, our model not only accurately captured the experimental measurements (Figure 3(h) in the manuscript and \Cref{figs5}) but also successfully predicted previously unforeseen results that were later confirmed experimentally (Figure 3(i,j,k) in the manuscript).

\Cref{figs9}(c) shows the force–deformation response when the model is partially unloaded, causing $n - m$ sliders to unlock $(1 < m < n)$. Upon reloading, these unlocked sliders relock initially and then unlock sequentially, closing the hysteresis sub-loop and exhibiting return point memory (RPM). Since the sequence of slider unlocking is identical during partial unloading and reloading, the hysteresis sub-loop is expected to follow Masing behavior. In \Cref{figs9}(d), we used the DDSSF model to compare the partially unloaded portion of the sub-loop with the reloading portion for various amounts of unloading. For smaller unloading amplitudes, the hysteresis sub-loops approximately follow Masing behavior, whereas for larger unloading, they deviate—consistent with the experimental observations (Figure 1(h) in the manuscript and \Cref{figs3}(d)).

\subsubsection*{DDSSF model in the continuum limit}

The fitted stiffness and friction coefficient values may be influenced by the level of discretization applied to the experimental unloading curve. A finer discretization leads to a greater number of spring-slider pairs in the model, resulting in a smoother fit. For simpler loading cases, a small number of spring-slider pairs may be sufficient to accurately capture the experimental response (e.g., $n=25$ in \Cref{figs9}(b)). However, for more complex loading scenarios involving multiple partial unloadings and reloadings, additional spring-slider pairs are necessary to detect subtle changes in the input force (e.g., $n=826$ in \Cref{figs9}(c,d)). As these pairs are connected in series, the fitted stiffness of each individual spring increases with the number of pairs $(n)$. The extent of further discretization is constrained by the number of data points in the experimental curve. Nevertheless, quasistatic data collected using commercial load frames typically exhibit a high sampling rate. In \Cref{figs9}(b), a smaller number of data points are presented in the experimental curve to reduce clutter, even though the actual experimental sampling rate was much higher.

Alternatively, the unloading curve can be fitted with a smooth function, $F_T = f(x)$, over the desired region (e.g., $x > 0.2$ in \Cref{figs9}(b)), to obtain continuous values of $k$ and $\mu$. At an arbitrary point on this fitted curve, $(x, f(x))$, the expression for $\mu$ (\Cref{mum}) can be rewritten as follow

\begin{equation}
    \mu = \frac{F_{T,max}-f(x)}{F_{T,max}+f(x)} \;\;,\;\;\Delta F = F_{T,max}-f(x)
    \label{mucont}
\end{equation}

Similarly, \Cref{km} can be rewritten for a continuous unloading curve as follows

\begin{equation}
    d\left( \frac{dx}{df(x)} \right) = \frac{1}{k(1-\mu)}  
\end{equation}
\begin{equation}
    \left( \frac{d^2x}{df(x)^2} \right) df(x)= \frac{1}{k(1-\mu)}  
    \label{kcont}
\end{equation}

Here, we rewrite \Cref{mucont} as
\begin{equation}
    \frac{1}{\mu+1} = \frac{F_{T,max}+f(x)}{2F_{T,max}}
\end{equation}
On differentiating both sides, we get
\begin{equation}
    df(x) = -d\mu\times\frac{2F_{T,max}}{(\mu+1)^2}
    \label{dphi}
\end{equation}

Substituting \Cref{dphi} in \Cref{kcont}

\begin{equation}
    \left( \frac{d^2x}{df(x)^2} \right) \times d\mu\times\frac{2F_{T,max}}{(\mu+1)^2}= -\frac{1}{k(1-\mu)}  
    \label{kcont2}
\end{equation}

Substituting \Cref{mucont} in \Cref{kcont2} and rearranging the terms

\begin{equation}
    kd\mu = -\frac{F_{T,max}}{f(x)\times(F_{T,max}+f(x))}\times \frac{1}{d^2x/df(x)^2}
\end{equation}

\begin{equation}
    \kappa = -\frac{F_{T,max}}{f(x)\times(F_{T,max}+f(x))}\times \frac{1}{d^2x/df(x)^2}
    \label{kappa}
\end{equation}

where, $\kappa$ represents a continuous stiffness that remains unaffected by the level of discretization of the unloading curve, unlike the stiffnesses of discrete springs $(k_i)$, which increase with $n$. The negative sign in \Cref{kappa} cancels the negative sign of $d^2x/df(x)^2$, which arises from the concave-up nature of the unloading curve, resulting in a positive value of $\kappa$. Although $\kappa$ is not the stiffness of any individual spring, for an unloading curve described by a smooth function $f(x)$, it can be computed as a function of $x$ using \Cref{kappa}. The stiffnesses of the individual springs in a discrete DDSSF model can then be estimated by dividing $\kappa$ by $d\mu$, the difference in $\mu$ values between consecutive Coulomb sliders.

\subsection*{Loss factor and loss tangent}

Dynamic mechanical analysis (DMA) experiment is performed by applying multi-cycle sinusoidal deformation, with specified amplitude $(\epsilon_d)$ and frequency $(\omega)$, to a statically precompressed sample $(\epsilon_s)$, resulting in the following total strain

\begin{equation}
    \epsilon_T = \epsilon_s +\epsilon_d \sin(\omega t)
\end{equation}

For a linear viscoelastic material with a time-dependent relaxation modulus $E(t)$, the stress response consists of a transient component and a sinusoidal steady-state component, given by

\begin{equation}
    \sigma_T = \epsilon_sE(t)+\sigma_d\sin(\omega t+\delta)
\end{equation}
\begin{equation}
    \sigma_T-\epsilon_sE(t)=\sigma_d \sin(\omega t)\cos(\delta) + \sigma_d \cos(\omega t) \sin(\delta)
    \label{stex}
\end{equation}

Generally, after applying $\epsilon_s$, the material is allowed to relax until the transient component of the stress relaxes, after which a sinusoidal strain is applied. The sinusoidal component of the stress signal leads the strain by a phase angle $\delta$. Plotting the sinusoidal stress against the sinusoidal strain yields an elliptical hysteresis loop. A larger phase angle $\delta$ results in a broader loop. The average slope of the hysteresis loop gives the dynamic modulus $E_d$, which is defined as follows \cite{lakes2009viscoelastic}

\begin{equation}
    E_d = \frac{\sigma_d}{\epsilon_d}
    \label{Ed0}
\end{equation}

In experiments, $E_d$ can be determined by dividing the measured stress amplitude $(\sigma_d)$ by the applied strain amplitude $(\epsilon_d)$, as given by the equation above (\Cref{Ed0}). Dividing \Cref{stex} by $\epsilon_d$ yields

\begin{equation}
    (\sigma_T-\epsilon_sE(t))/\epsilon_d=E_s \sin(\omega t) + E_l \cos(\omega t)
\end{equation}

where $E_s = E_d \cos(\delta)$ and $E_l = E_d \sin(\delta)$ are the storage and loss moduli, respectively. The storage modulus is associated to the energy stored, while the loss modulus is associated to the energy dissipated. The area enclosed by the hysteresis loop, representing the energy dissipated per unit volume $(W_{{dis}})$, can be calculated as follows

\begin{equation}
    W_{dis} = \int_0^{2\pi/\omega} \sigma_d\sin(\omega t+\delta) \epsilon_d \cos(\omega t)\omega \;dt
\end{equation}
\begin{equation}
    W_{dis} = \pi E_l \epsilon_d^2
    \label{wdis}
\end{equation}

In experiments, $E_l$ can be calculated from the measured energy dissipated per unit volume, $W_{{dis}}$, and the applied dynamic strain amplitude, $\epsilon_d$, using the equation above. The storage modulus $E_s$ can be calculated using the following equation

\begin{equation}
    E_s=\sqrt{E_d^2-E_l^2}
    \label{Esto}
\end{equation}

For VACNT foams, we calculated the dynamic modulus from experimental measurements using \Cref{Ed0}. In contrast to viscoelastic materials, the stress and strain signals for VACNT foams are not phase shifted (Figure 2(k) in manuscript). The energy dissipation in VACNT foams arises not from viscoelasticity that introduces phase difference but from nanoscale friction, resulting in a hysteresis loop that is biconvex in shape rather than elliptical. As suggested by \Cref{wdis}, the magnitude of the loss modulus is directly related to energy dissipation. Accordingly, we define an effective loss modulus for VACNT foams using \Cref{wdis}, while the effective storage modulus is calculated using \Cref{Esto}.
Nevertheless, the dynamic modulus of VACNT foams cannot be represented in the complex form $(E_d^* = E_s + iE_l)$, as is commonly done for viscoelastic materials \cite{lakes2009viscoelastic}. The ratio of loss modulus to storage modulus gives the tangent of $\delta$, also known as the loss tangent ($tan(\delta)$). For VACNTs, we define this ratio as the loss factor $\zeta = {E_l}/{E_s}$, which quantifies the damping ratio.

\subsection*{Group velocity of 1D monoatomic layered chain}

\begin{figure}[H]
	\centering
	\includegraphics[width=\textwidth]{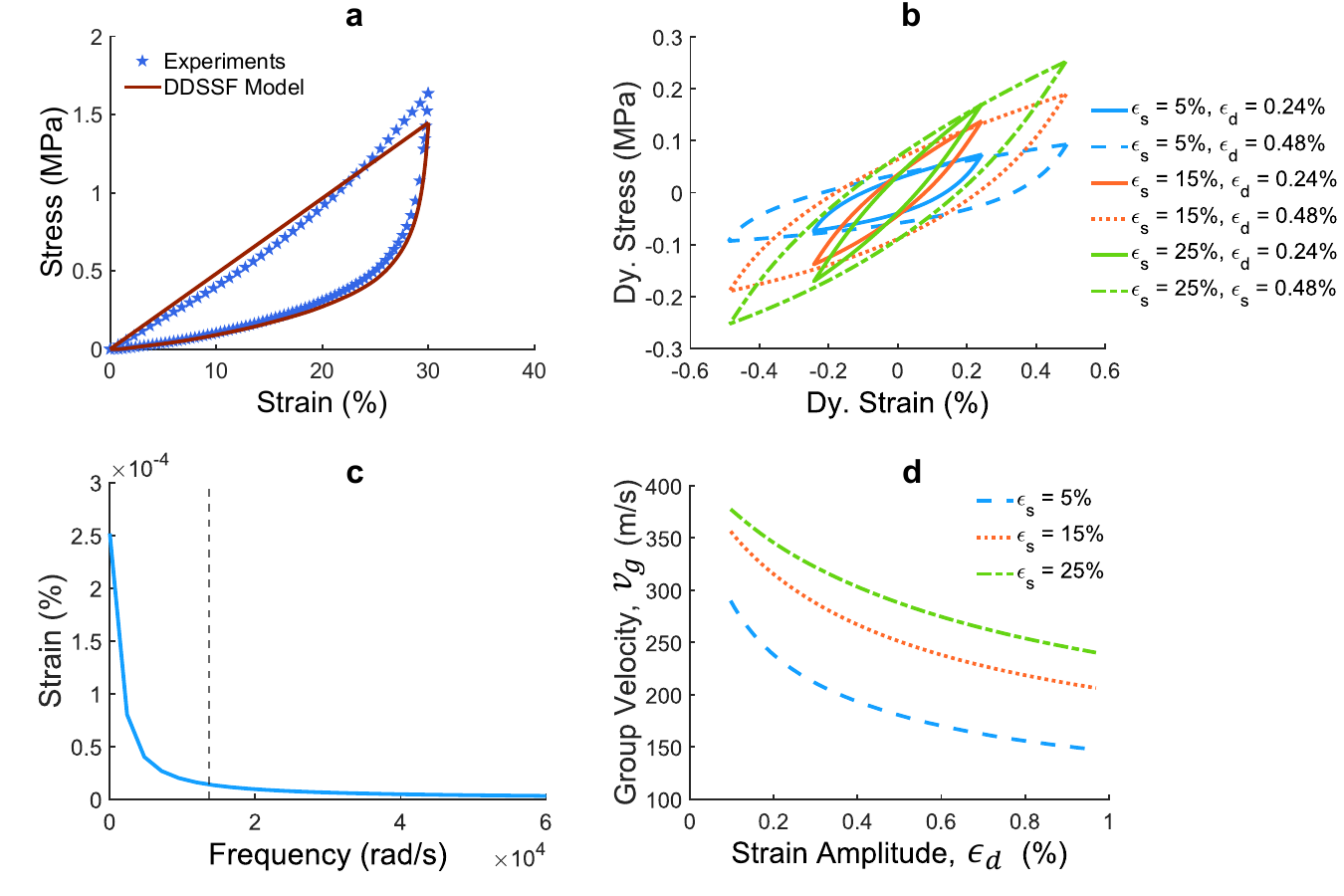}
	\caption{ \textbf{Group velocity estimation.} (a) DDSSF model fit on the experimentally measured average stress-strain response of $17$ samples used in wave propagation experiment. (b) Dynamic hysteretic subloops for various $\epsilon_d$ and $\epsilon_s$ calculated using DDSSF model. (c) Frequency spectrum of strain signal measured in experiments. (d) Group velocity as a function of $\epsilon_d$ and $\epsilon_s$}
	\label{figs10}
\end{figure}

In the manuscript, we described our experimental measurements of the amplitude and precompression-dependent speed of a traveling stress pulse in a chain of VACNTs with aluminum interlayers (Figure 4 in manuscript). Here, using the DDSSF model, we reproduce similar behavior in the group velocity of a pulse traveling through a one-dimensional phononic crystal, where VACNTs are modeled as elastic springs and aluminum cylinders as rigid masses. \Cref{figs10}(a) shows the average preconditioned stress-strain response of the 17 VACNT samples used in the experiments. The stress-strain global hysteresis is fitted with the DDSSF model ($n = 40600$), following the method described earlier (\Cref{figs9}(b)). Using the fitted stiffnesses and friction coefficients, we then estimate the dynamic modulus as a function of $\epsilon_d$ and $\epsilon_s$ by simulating the hysteretic sub-loop response (\Cref{figs10}(b)) with our MATLAB script (\hyperref[section:sd4]{GitHub}). 

Let $A$ denote the cross-sectional area and $h$ the average height of the VACNT samples, the dynamic stiffness can be defined as follows

\begin{equation}
    k_d = E_d\frac{A}{h}
\end{equation}

Assuming $m$ is the mass of the aluminum interlayer cylinders with height $h_l$ and cross-sectional area $A_l$, a characteristic frequency of the system can be defined as follows

\begin{equation}
    \omega_0=\sqrt{\frac{k_d}{m}}=\sqrt{\frac{E_dA}{mh}}
    \label{omeg0}
\end{equation}

For a 1D phononic mass-spring chain, the dispersion relation is given by \cite{deymier2013acoustic}

\begin{equation}
    {\omega}= 2{\omega_0} \left| \sin\left( \frac{\nu d}{2}\right) \right|
\end{equation}

where $\nu$ is the wave number and $d$ is the characteristic length of the unit cell, i.e., $h + h_l$. The expression for the group velocity is obtained by differentiating the dispersion relation, as follows

\begin{equation}
    v_g=\frac{\partial \omega}{\partial \nu} =  \omega_0 d\left| \cos\left ( \frac{\nu d}{2} \right) \right|
    \label{grp}
\end{equation}

In \Cref{figs10}(c), we plot a representative frequency spectrum obtained by applying Fourier transform to the time-domain strain signal measured in experiments (see \Cref{figs6}(b)). The dashed black line indicates $\omega_0 = 13613\;rad/s$, calculated using \Cref{omeg0} by inputting $E_d$ corresponding to $\epsilon_s = 5\%$ and $\epsilon_d = 0.96\%$, which yields the lowest possible value of $\omega_0$ within the range of $\epsilon_s$ and $\epsilon_d$ considered (\Cref{figs10}(d)).
However, the dominant frequencies in the spectrum are significantly lower than $\omega_0$ (\Cref{figs10}(c)). Therefore, we assume $\omega \ll \omega_0$, which allows us to approximate $\cos(\nu d/2) \approx 1$. Under these assumptions, we rewrite \Cref{grp} as follows

\begin{equation}
        v_g=\frac{\partial \omega}{\partial \nu} =  \omega_0 d
\end{equation}

Substituting $\omega_0$ from \Cref{omeg0}

\begin{equation}
        v_g =  d\sqrt{\frac{E_dA}{mh}}
\end{equation}

By substituting $E_d$---measured using the DDSSF model as a function of $\epsilon_d$ and $\epsilon_s$---along with $A = 19.64\;{mm}^2$, $h = 2.06\;{mm}$, $m = 0.61\;{g}$, and $d = 10.06\;{mm}$, we obtain the theoretically estimated group velocity $(v_g)$, which is plotted as a function of $\epsilon_s$ and $\epsilon_d$ in \Cref{figs10}(d). Both the magnitude and the trend of the estimated group velocity clearly align with the experimental measurements, demonstrating the accuracy of our theoretical approach and DDSSF model. The $x$-axis in \Cref{figs10}(d) is $\epsilon_d$, which quantifies the intensity of the input pulse, as captured by $v_i$ in the experimental data plot (Figure 4(b) in the manuscript). Establishing a theoretical relationship between $\epsilon_d$ and $v_i$ would require a more detailed analysis, which is beyond the scope of this work.


\subsection*{Amplitude-dependent wave modulation using tailorable phase gradients}

We arrayed five parallel chains of VACNT–aluminum interlayers and replicated this pattern in six identical radial sectors to build a hexagonal waveguide (see Figure 4(c) in the manuscript). Because the overall response of the structure is the superposition of these sectors, we may analyze just one sector and its five chains. The net output of the hexagonal waveguide then follows directly. 

Each chain here is separated by an equal spacing $s$, to create a periodic waveguide based-phased array and a uniform input pulse of negligible pulse width was given. This system exhibits two distinct wave propagation phenomena. First, there is the uniaxial wave propagation along each chain of the periodic structure \cite{brillouin1946wave}, where dispersion arises from the inherent periodicity and attenuation occurs due to frictional dissipation within the VACNT foams. Second, once the waves exit the periodic structure, they behave as localized dynamic loads that excite elastic spherical waves in the semi-infinite receiver medium \cite{meyers1994dynamic}, which is assumed to follow a homogeneous isotropic linear elastic (HILE) response. To rigorously analyze the system, one could first solve the uniaxial wave equation for the periodic structure—using, for example, a transfer matrix method that treats the repetitive unit cell as a continuum with a frequency-independent complex modulus, accounting for frequency independent dissipation of VACNT foam—and then apply the resulting field as the initial condition for the spherical wave equation (\Cref{eqsphericalwave}) in the receiver medium. The well-established analytical solution for outward spherical wave propagation in a HILE medium (\Cref{eqoutwardspherical}) would then provide the stress (pressure) and deformation fields within the receiver medium.

\begin{equation}
\frac{\partial^{2} u}{\partial t^{2}}
\;=\;
c^{2}\!\left(
      \frac{\partial^{2} u}{\partial r^{2}}
      + \frac{2}{r}\,\frac{\partial u}{\partial r}
    \right)
    \label{eqsphericalwave}
\end{equation}

\begin{equation}
u(r,t) \;=\; \frac{1}{r}\,F\!\bigl(r - c\,t\bigr)
\label{eqoutwardspherical}
\end{equation}

\begin{figure}[H]
	\centering
	\includegraphics[width=\textwidth]{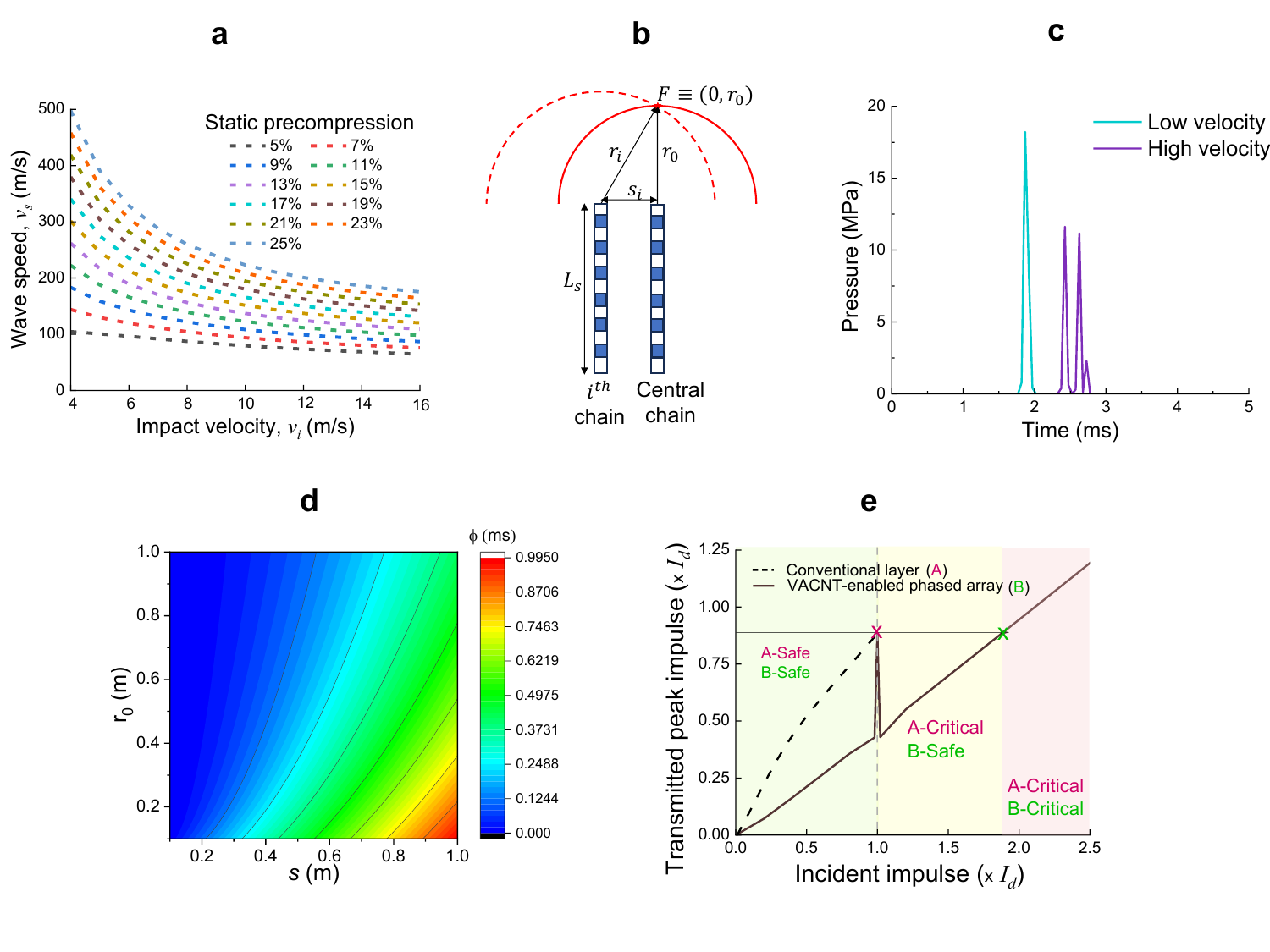}
	\caption{ \textbf{Analytical modeling of phased gradient-based wave modulation approach.} (a) Instance of coalescing of the spherical evolving from the $i^{th}$ chain with the that of the central chain  (b) Interpolated wave speed distribution in the periodic wave guide (c) Pressure outputs obtained at $\mathcal{F}$ for low velocity impact and high velocity impact scenarios (d) Design contour for the selection of appropriate spacing and phase delay based on the focal location (e) Transmitted peak impulse at a given instance vs incident impulse for a conventional protective layer and VACNT-enabled phased array system}
	\label{figs11}
\end{figure}

Although this approach would provide a more accurate estimation of the pressure in the receiver medium, our primary objective is to illustrate that the experimentally demonstrated robust wave speed tuning—using dynamic amplitude and precompression—can be strategically exploited to engineer VACNT-enabled phased array acoustic wave limiters. Therefore, we performed a simplified analysis for the wave propagation through the periodic chains by directly utilizing our experimental results to interpolate a wave speed distribution as a function of impact velocity and precompression (\Cref{figs11}(a)). For the spherical wave propagation in the receiver medium---with an elastic wave speed $(c_R )$ of 1000 $m/s$---we directly applied the analytical solution (\Cref{eqsphericalwave}). For the illustration, we assumed the same system dimensions ($L$ as the distance between strain gages), applied a 5\% precompression to the central chain, 7\% to the adjacent chains, and 25\% to the outer chains, and selected lower and higher impact velocities of 5 $m/s$ and 6 $m/s$, respectively. Note that the stress pulses at the interface are attenuated due to the dissipation within the VACNT foams in a complex manner that would normally require detailed analytical treatment (using DDSSF model). However, we assumed an average loss factor of 0.5---yielding a 50\% attenuation of the stress pulses at the interface---which is a reasonable estimate for VACNT foams. 
This initial static precompression setting leads the wave to travel slowest in the central chain and increasingly faster in those farther from the central one, resulting in a leading phase ($\phi_i$) value for the chains farther from the central one with respect to the central one. (The phase of the central chain is hence $0$).  The average wave speeds in the individual periodic chains $c_i$ were calculated using the wave speed distribution shown in the figure. The leading phase, $\phi_i$ of the $i^{th}$ chain from the central one can be calculated using \Cref{eqphasedelay}.

\begin{equation}
\phi_i \;=\; L\left(\frac{1}{c_0} - \frac{1}{c_i}\right)
\label{eqphasedelay}
\end{equation}

Where, $c_0$ and $c_i$ is the average wave speeds of the pulse through the central chain and the $i^{th}$ chain from the central chain, respectively.

The radius of the spherical wave fronts due to the pulse from the $i^{th}$ chain—from the central chain—at the coalescence instance can be found using the geometric expression, (see \Cref{figs11}(b))

\begin{equation}
r_i \;=\; \sqrt{\,r_0^{2} + s_i^{2}\,}
\label{eqr_i}
\end{equation}

where, $r_0$ is the radius of the spherical wave front corresponding to the central chain at the instance of coalescing and is the $s_i$ is the horizontal spacing between the central chain and the $i^{th}$ chain from it. Meanwhile, an incremental propagation relation for $r_i$ can be written in terms of $r_0$, $c_R$, and $\phi_i$ as follows,

\begin{equation}
r_i \;=\; r_0 + c_R\,\phi_i
\label{eqri_2}
\end{equation}

By substituting \Cref{eqr_i} in \Cref{eqri_2}, the spacing $s_i$ required for the coalescence to take place at $F\equiv(0,r_0)$ can be found in terms of $r_0$ from the following coalescence relation. 

\begin{equation}
s_i \;=\; \sqrt{\,2\,c_R\,\phi_i\,r_0 \;+\; c_R^{2}\,\phi_i^{2}\,}
\label{eqspacing}
\end{equation}

Because the five parallel chains are equally spaced, we can apply the coalescence relations to both the chains—immediately adjacent to the central chain and those at the far ends—yielding two simultaneous relations. Solving these equations gives a unique solution for $s$ and $r_0$ (0.64 $m$ and 0.98 $m$, respectively) for coalescence under the selected precompression arrangement in the low-amplitude scenario.
However, in the other scenario, increasing the input amplitude disrupted this coalescence, causing the wave fronts to arrive asynchronously and resulting in a more attenuated, pulse-stretched response; notably, the higher amplitude scenario produced an output amplitude approximately 1.6 times lower than that of the smaller amplitude. Because VACNT foam dissipation can increase with amplitude (Figure 2(j) in the manuscript), a more rigorous analysis would likely accentuate this counterintuitive input-output trend. \Cref{figs11}(c) presents the output responses obtained at the predetermined coalescing point $\mathcal{F}$ by design for the considered low velocity and high velocity impact scenarios. Finally, to maintain focus on qualitative behavior, normalized time units have been used for the plots presented in the manuscript.

Additionally, we propose a systematic approach to design amplitude-dependent wave limiters that enable coalescence at a targeted input amplitude while attenuating and pulse-stretching any excess amplitudes. For a given target amplitude, a suitable peak output—representing a safely transmittable level—can be selected. This coalescence output amplitude is determined by several factors: the loss factor ( $\zeta$) of the VACNT foams used in the phased array, the coalescence location ($r_0$), and the phase of each chain ($\phi_i$). Neglecting the phase contributions initially, one can choose VACNT foams with an appropriate $\zeta$ and determine the optimal $r_0$. Then, using the provided contour plot (\Cref{figs11}(d)), an appropriate combination of spacing ($s_i$) and $\phi_i$ for each chain—from the central chain outward—can be selected to yield the required wave speeds in each chain. Ultimately, the resulting wave speed ($v_s$)  distribution as a function of dynamic force and precompression is used to derive the necessary precompression in each chain, thereby achieving the desired phase shifts and, in turn, the target coalescence outcome.

In impact-mitigation technologies, this VACNT-enabled phased array system offers a decisive advantage over conventional protective layers that rely solely on material damping and dispersion to keep transmitted impulse below a critical limit. Specifically, in those traditional designs, the incident stress pulse traverses the protective layer as a single, coherent wave front whose amplitude diminishes only as energy is dissipated due to material constitutive behavior. By contrast, our system deliberately splits the incoming impulse into several wave fronts due to carefully engineered phase offsets. These fronts are tuned to coalesce constructively only for a pre-selected reference impulse, producing a peak pressure just shy of the transmittable critical threshold. Should a stronger impact occur, the phase-shifted fronts no longer align in space and time; instead, they arrive out of synchrony, spreading the hazardous impact energy over a larger area and a longer duration. This spatiotemporal redistribution offers an additional mean of attenuation, effectively widening the safe-operating window and allowing the structure to tolerate impulses that would overwhelm conventional protective systems. 

We quantified this advantage with the analytical model that compares transmitted peak impulse at a given instance versus incident impulse for a conventional protective layer and our phased-array system. We assigned the conventional layer the same loss factor as the VACNTs to isolate the effect of phase tailoring. As shown in \Cref{figs11}(e), for VACNT-based array, the peak transmitted impulse occurs at the instance of coalescence however, it enables an additional safe-operating region (shaded in yellow) beyond the design incident impulse $I_d$ , demonstrating the extra protection achieved by our strategy.

\newpage
\section*{Supplementary Figures}

\newpage


\end{document}